\documentclass[printer]{aa}
\usepackage{amssymb}
\usepackage{amsmath}
\usepackage{txfonts}
\usepackage{graphicx}
\usepackage{natbib}
\usepackage{rotating}
\usepackage{soul}
\usepackage{url}
\usepackage{epsfig}
\usepackage{longtable}

\usepackage{color}
\usepackage[utf8]{inputenc}
\usepackage{url}

\usepackage{bbold}

\begin{document}

\title{Reconstruction of the two-dimensional gravitational potential of galaxy clusters from X-ray and Sunyaev-Zel'dovich measurements}
\titlerunning{SZ \& X-rays potential potential reconstruction on a grid}

\author{C. Tchernin\inst{1,2}, M. Bartelmann\inst{1}, K. Huber\inst{3,4}, A. Dekel\inst{5,6}, G. Hurier\inst{7}, C. L. Majer\inst{1}, S. Meyer\inst{1}, E. Zinger\inst{5}, D. Eckert\inst{8}, M.~Meneghetti\inst{9,10},  J. Merten\inst{11}}
\institute{\inst{1}Universität Heidelberg, Zentrum für Astronomie, Institut für Theoretische Astrophysik, Philosophenweg 12, 69120 Heidelberg,\\ e-mail: Tchernin@uni-heidelberg.de, 
\\\inst{2}Department of Astronomy, University of Geneva, ch. d'Ecogia 16, 1290 Versoix, Switzerland,
\\\inst{3}Excellence Cluster Universe, LMU, Boltzmannstr. 2, 85748 Garching, Germany,
\\\inst{4}Ludwig-Maximilians-Universität, Fakultät für Physik, Universitäts-Sternwarte, Scheinerstraße 1, 81679 München, Germany,
\\\inst{5}Center for Astrophysics and Planetary Science, Racah Institute of Physics, The Hebrew University, Jerusalem 91904, Israel,
\\\inst{6}SCIPP, University of California, Santa Cruz, CA, 95064, USA,
\\\inst{7}Centro de Estudios de Fisica del Cosmos de Arago (CEFCA),Plaza de San Juan, 1, planta 2, E-44001, Teruel, Spain,
\\\inst{8}Max-Planck Institute for Extraterrestrial Physics (MPE), Giessenbachstrasse 1, 85748 Garching, Germany,
\\\inst{9}Osservatorio Astronomico di Bologna, INAF, via Ranzani 1, 40127, Bologna, Italy, 
\\\inst{10}INFN, Sezione di Bologna, viale Berti Pichat 6/2,40127, Bologna, Italy,
\\\inst{11}Department of Physics, University of Oxford, Denys Wilkinson Building, Keble Road, Oxford OX1 3RH, U.K.}

\authorrunning{C. Tchernin et al.}


\abstract
{The mass of galaxy clusters is not a direct observable, nonetheless it is commonly used to probe cosmological models. Based on the combination of all main cluster observables, that is,\ the X-ray emission, the thermal Sunyaev-Zel'dovich (SZ) signal, the velocity dispersion of the cluster galaxies, and gravitational lensing, the gravitational potential of galaxy clusters can be jointly reconstructed.}
{We derive the two main ingredients required for this joint reconstruction: the potentials individually reconstructed from the observables and their covariance matrices, which act as a weight in the joint reconstruction. We show here the method to derive these quantities. The result of the joint reconstruction applied to a real cluster will be discussed in a forthcoming paper.}
{We apply the Richardson-Lucy deprojection algorithm to data on a two-dimensional (2D) grid. We first test the 2D deprojection algorithm on a $\beta$-profile. Assuming hydrostatic equilibrium, we further reconstruct the gravitational potential of a simulated galaxy cluster based on synthetic SZ and X-ray data. We then reconstruct the projected gravitational potential of the massive and dynamically active cluster Abell 2142, based on the X-ray observations collected with \textit{XMM-Newton} and the SZ observations from the \textit{Planck} satellite. Finally, we compute the covariance matrix of the projected reconstructed potential of the cluster Abell 2142 based on the X-ray measurements collected with \textit{XMM-Newton}.}
{The gravitational potentials of the simulated cluster recovered from synthetic X-ray and SZ data are consistent, even though the potential reconstructed from X-rays shows larger deviations from the true potential. Regarding Abell 2142, the projected gravitational cluster potentials recovered from SZ and X-ray data reproduce well the projected potential inferred from gravitational-lensing observations. We also observe that the covariance matrix of the potential for Abell 2142 reconstructed from \textit{XMM-Newton} data sensitively depends on the resolution of the deprojected grid and on the smoothing scale used in the deprojection.}{ We show that the Richardson-Lucy deprojection method can be effectively applied on a grid and that the projected potential is well recovered from real and simulated data based on X-ray and SZ signal. The comparison between the reconstructed potentials from the different observables provides additional information on the validity of the assumptions as function of the projected radius.  }
\maketitle

\section {Introduction}
As the most massive gravitationally bound systems in the Universe, galaxy clusters provide information on the evolution of the cosmic large-scale structures. The statistical properties of the galaxy cluster population can be described as a function of the clusters' mass and redshift. The comparison to the mass function, that is,\ the number density of dark-matter halos as a function of the redshift, whose high-mass tail is highly sensitive to the cosmological model \citep[e.g.,][]{press74, tinker08,watson13}, is commonly used to set cosmological constraints. Yet, cluster masses are not directly observable, hence scaling relations are needed to convert the observables (e.g.,\ the gas temperature, the X-ray luminosity, or the richness) to an estimate of cluster mass \citep[see for instance,][]{pratt09,andreon10}. Also, for the most relaxed clusters and in the regions of the cluster expected to be in hydrostatic equilibrium, the hydrostatic mass derived from X-ray measurements shows around 10\% scatter between the true cluster mass and its estimate \citep{app16}. Based on such mass estimates, observations typically confirm the $\Lambda$CDM cosmology \citep[e.g.,][]{planck15}. However, the scatter around the true mass may mask subtle deviations from this fiducial model.

Here, we aim to derive a complementary method to characterize galaxy clusters using their gravitational potentials rather than their masses. Whether or not any cosmological constraints conventionally derived from the mass function would benefit from replacing the mass with the gravitational potential is the main question that the project, which this paper is a part of, is addressing. The expected advantage of the potential with respect to the mass is that the potential can be derived from lensing information without requiring the integration over an unknown boundary surface and it is therefore closer to the observations than the cluster mass. Thus, this may improve the sensitivity of the statistical methods based on the galaxy clusters as cosmological probes.\\ The goal of this paper is to describe the method to reconstruct a cluster potential from multiwavelength observations. Upcoming papers will study in detail the systematic errors related to the assumptions of the potential reconstruction method \citep{tchernin18} and to the jointly reconstructed potential \citep{huber18}.

Observations of galaxy clusters are related to the gravitational potential as follows: Velocity dispersions of the cluster member galaxies measure the potential gradient, gravitational lensing, the potential curvature, and the intra-cluster medium (ICM) emitting the X-ray, and the Sunyaev-Zel'dovich (SZ) signals directly trace the gravitational potential in hydrostatic equilibrium \citep[see e.g.,][for a review]{limousin13}. Furthermore, each of these observables probes different scales: while the X-ray and strong-lensing signals emerge from the cluster center, the weak-lensing and thermal SZ signals trace the potential further out. Combining observables, we expect to recover cluster potentials on a wider range of radii \citep[e.g.,][]{planck13,limousin13}. 

To reconstruct cluster potentials, we use the Richardson-Lucy deprojection algorithm \citep[R-L,][]{lucy74,lucy94}. The one-dimensional (1D) R-L method has been successfully tested against other deprojection schemes \citep[see for instance][where the outcome of the R-L deprojection was compared to the onion-peeling deprojection method]{tchernin15}. The R-L scheme offers the substantial advantage of allowing the deprojection of all lines-of-sight separately, and thus it is applicable to incomplete data sets, allowing us to exclude from the analysis data coming from regions where the assumptions made may not hold (such as, at the cluster center or in the outskirts). This is an important feature of this method, as discussed in Sect.~\ref{sec:discussion}.

So far, cluster potentials have been successfully reconstructed from cluster observables in one dimension by treating each observable separately: the SZ effect \citep{majer13}, the galaxy kinematics \citep{sarli14} and the X-ray signal \citep{konrad13}. A complete joint analysis is however still missing.

In this paper, we outline our method to derive the jointly constrained two-dimensional (2D) gravitational potential within the SaWLens framework \citep[see e.g,][]{merten14}. We first generalize the R-L algorithm to a 2D grid, which will allow us to perform the joint analysis of the cluster potential pixel per pixel. Then, to perform the joint analysis, we derive the covariance matrices of the reconstructed potentials. The latter enters directly into the fitting procedure as a relative weight for the different contributions.

The structure of the paper is as follows: In Sect.\ \ref{sec:deproj_method}, we show the different steps leading from 1D to 2D R-L deprojection in the spherically symmetric case. In Sect.\ \ref{sec:deproj_proof}, we test the deprojection algorithm on a simulated $\beta$-profile emission map. We apply the method to unbinned data (\ref{sec:betaprof_sansvor}) and to data binned into a Voronoi tessellation (\ref{sec:betaprofvor}). The gravitational potential reconstruction from X-ray and SZ observations is outlined in Sect.\ \ref{sec:recons_method}. In Sect.\ \ref{sec:recons_g1}, we reconstruct the projected gravitational potential of a simulated cluster based on simulated X-ray and SZ data, and compare it to the true projected gravitational potential. In Sect.\ \ref{sec:recons_A2142} we reconstruct the projected gravitational potential of the cluster Abell 2142, based on real X-ray observations with \textit{XMM-Newton}, and SZ observations with the \textit{Planck} satellite. In Sect.\ \ref{sec:join} we outline the method for the joint potential reconstruction, and we compute the covariance matrix of the projected reconstructed potential of the cluster Abell 2142 based on its X-ray observations with \textit{XMM-Newton}.  We then discuss and conclude in Sects.\ \ref{sec:discussion} and \ref{sec:conclusion}.

\section {Deprojection on a 2D grid}\label{sec:deproj}
\subsection {The R-L deprojection Method}\label{sec:deproj_method}\label{sec:methodRL}
\subsubsection{Generalization of the 1D R-L to a grid}
In this section, we review the important steps allowing us to pass from the one- to the two-dimensional deprojection. We refer the reader to the papers of \citet{lucy74,lucy94} for details on the R-L method and to \citet[][]{konrad13} for its application to the 1D case.

Assuming spherical symmetry, the 1D spherical kernel is \citep{konrad13}
\begin{equation}\label{eq:sphk}
  K(s|r)=\frac{1}{N(r)}\frac{\ r }{\sqrt{r^2-s^2}}\Theta(r^2-s^2)\;,
\end{equation}
where $s$ is the projected and $r$ the three-dimensional (3D) radius, with $s$ and $r$ being related to the line-of-sight coordinate $z$ by $r^2=s^2+z^2$. $\Theta$ is the Heaviside step function and $N(r)$ a radially dependent normalization constant. The 2D quantity $g(s)$ can thus be obtained by projecting the 3D function $f(r)$ along the line-of-sight\;,
\begin{equation}
  g(s)=\int dr K(s|r) f(r)\;.
\end{equation}
To generalize these two equations to two dimensions, we define the two axes $s_1$ and $s_2$ and set the origin of the coordinate system at the center of the grid. To avoid difficulties with odd numbers of pixels, we assign the coordinates to pixel centers.
\\The generalized 2D spherical kernel is then similar to Eq.~(\ref{eq:sphk}), but with $s^2(i,j)=s_2^2(i)+s_1^2(j)$, with $(i,j)$ being the indices of the matrix representing the grid.

The spherical kernel needs to be normalized on the grid,
\begin{equation}\label{eq:Nk}
  \int ds_1\int ds_2 K(s|r)=1\;.
\end{equation}
If $s_1$ and $s_2$ are continuous quantities, this condition is satisfied for $N=2\pi r^2$.

We note that the functions $f$ and $g$ need to be normalized with respect to the integrals over their domains, therefore, all the quantities shown in this paper are normalized to 1. The correct units can nevertheless be tracked through the deprojection process and correctly recovered. However this is not needed for the scheme of this paper.

\subsubsection{Technical detail}

\begin{figure}
\begin{center}
  \includegraphics[height=\columnwidth,angle=270]{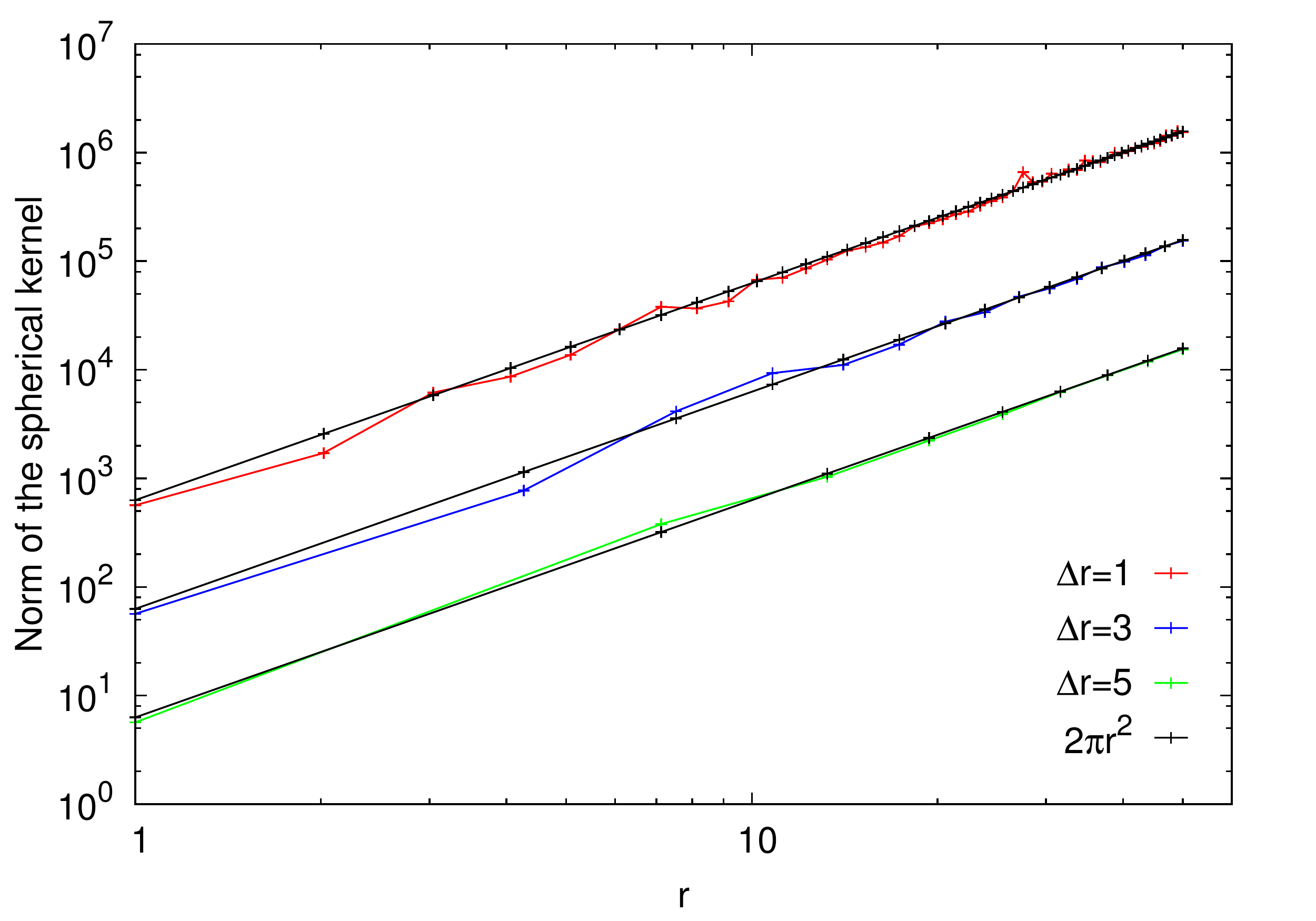}
\caption{Normalization of the spherical kernel (Eq.~(\ref{eq:Nk})) as a function of the 3D radius $r$ for three choices of binning in $r$ and discretized values of $s_1$ and $s_2$. In red for a linear binning of $\Delta r=1$; in blue for a linear binning of $\Delta r=3$ and in green for a linear binning of $\Delta r=5$. These normalizations are compared to the results of Eq.~(\ref{eq:Nk}) in the case of continuous values of $s_1$ and $s_2$, in black. For better visibility, the normalizations for $\Delta r=1$ and $\Delta r=3$ have been multiplied by 100 and 10, respectively.}
\label{fig:Nk}
\end{center}
\end{figure}

\begin{itemize}
\item{}\textit{Choice of the binning in $r$:}

The discretization of $s_1$ and $s_2$ introduces discontinuities in the deprojected profile. Indeed, since the integral over the spherical kernel is performed over $s_1^2+s_2^2<r^2$, the discretization of $s_1$ and $s_2$ causes wiggles in the normalization ($N$ from Eqs.~(\ref{eq:sphk}) and ~(\ref{eq:Nk})). Their strength depends on the binning in the 3D radius $r$. This is illustrated in Fig.~\ref{fig:Nk}, where $N$, computed for different binnings in $r$ (we choose linear intervals in $r$ of $\Delta r=1$, 3 and 5 pixels) is plotted as a function of $r$. In order to illustrate the effects of the discretization of $s_1$ and $s_2$, we compared the result of Eq.~(\ref{eq:Nk}) for discrete and continuous values of $s_1$ and $s_2$ (for better readibility, we multiplied the results for $\Delta r=1$, 3 and 5 by 100, 10 and 1, respectively). As expected, the smaller the binning in $r$, the more wiggles appear. We thus need to choose bins in $r$ such as to minimize the effect of the discretization of $s_1$ and $s_2$ while retaining enough information for deprojecting the signal.

For this purpose, we choose to bin $r$ such that the quantity we want to deproject has an almost constant signal-to-noise ratio (S/N). The binning in $r$ needs to be optimized for each observation.\\

\item{}\textit{Upper limit for $r$:} 

Due to the assumed spherical symmetry, only the pixels of the grid satisfying $r>\sqrt{s_1^2+s_2^2}$ can be used for the deprojection (see Eq.~(\ref{eq:sphk})). The largest allowed value of $r$ equals the radius of the largest circle enclosed in the grid. This implies that the corners of the grid are ignored in the deprojection. We can see this effect for instance in Fig~\ref{fig:mapsansvor_deproj}.\\

\item{}\textit{Lower limit for $r$:} 

The minimal value of $r$ cannot be smaller than the side length of a pixel, which corresponds to the spacing between two consecutive values of $s_{1 (2)}$ on the grid. For the reprojection, we set all pixels at a distance smaller than the first value of $r$ ($r[0]$) to the value of the reprojected profile at $r[0]$. \\

\item{}\textit{Choice of the parameters of the R-L deprojection:} 

The R-L deprojection method depends on two parameters, both introduced to avoid the overfitting: the regularization parameter $\alpha$ and the smoothing scale $L$ \citep[see][for details]{lucy74,lucy94}. While both parameters should be adapted to the data, there is no general criterion to find the best value for $\alpha$ \citep[][]{lucy94}. An approach to estimating the optimal value of $\alpha$ for fixed $L$ can be found in \citet{majer16}. We emphasize, however, that the choice of $\alpha$ and $L$ influences the deprojection only very little for $\alpha>0$ and values of $L$ smaller than half of the size of the observed region. In the study presented here, we have chosen both parameters to
depend on the cluster-centric distance: we take $\alpha$ inversely proportional to the S/N, and set $L$ to increase linearly with the radius. This choice is motivated by the fact that high values of $\alpha$ give more weight to the regularization prior than to the data. This implies that the deprojected profile will carry the information of the regularization prior, which in our case is a Gaussian with smoothing scale $L$ \citep[see][for the actual form of the regularization term]{konrad13}. Therefore, the values of $\alpha$ and $L$ should increase with the distance from the cluster center, to be maximal where the statistical fluctuations are the largest. 
This implies that the values of these two parameters depend strongly on the binning in $r$. This is an optimized approach compared to the previous reconstruction studies \citep{konrad13,majer13,sarli14}, where the value of $\alpha$ was kept constant over the entire cluster.\\

\item{}\textit{Importance of the field-of-view of the observation used for the potential reconstruction:} 

The potential is recovered from the information contained in the observables only, and thus restricted to the domain of the data. For directly comparing the potentials recovered from two different probes, like from SZ and X-rays, we truncate the larger field-of-view here. However, we note that this truncation is not necessary in our joint analysis (see Sect. 6), where we combine constraints in the full SaWLens framework \citep[see e.g.,][]{merten14} to jointly reconstruct the potential using simultaneously information pixel by pixel. Therefore, gaps in the potential reconstruction due to lack of data will be excluded from the joint reconstruction.\\

\item{}\textit{Assumed symmetry}: 

In this paper, we only assume spherical symmetry. This ideal case is chosen because it leads us to an understanding of the covariance induced by the reconstruction method in the simplest scenario (discussed in detail in Sect. \ref{sec:join} below). This is clearly a first step prior to studying the more complex correlations expected in the spheroidal case (which is ongoing).
\end{itemize}

\subsection{Application to a grid with a $\beta$-profile signal}\label{sec:deproj_proof}
We apply the R-L deprojection to a $\beta$-profile simulated map, which is a good description of the ICM \citep[e.g.,][]{king}, and mimics the X-ray surface brightness \citep[e.g.,][]{jones84}. 
This allows us to test the deprojection method in the most ideal case. Then, increasing stepwise the difficulty of the reconstruction, we apply the potential reconstruction method to a simulated cluster (in Sect.~\ref{sec:recons_g1}) and to real data (in Sect.~\ref{sec:recons_A2142}).

We produce 100 realizations of the $\beta$-profile map by adding randomly sampled Poisson noise. The simulated maps have 100x100 pixels, and the center of the ``cluster'' is set at the center of the map (see the left panel of Fig.\ \ref{fig:mapsansvor_deproj}).
\\The $\beta$-profile as a function of $x$, the distance to the cluster center, can be expressed as
\begin{equation}\label{eq:betaprofile}
n(x)= n_0\cdot\left(1+\left(\frac{x}{r_c}\right)^2\right)^{(-3\beta+0.5)},
\end{equation}
where we choose $n_0=200$ counts, $\beta=0.67$ and $r_c=10$ pixels. 

\subsubsection{Deprojection without applying Voronoi tessellation}\label{sec:betaprof_sansvor}
\begin{figure}
\begin{center}
  \includegraphics[height=0.7\columnwidth,angle=0]{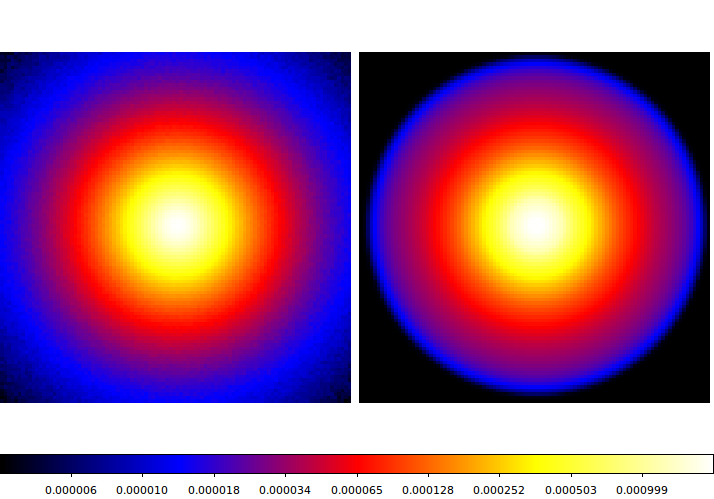}
\caption{Maps illustrating the result of the R-L deprojection algorithm on simulated $\beta$-profile maps (Eq.~(\ref{eq:betaprofile})) assuming spherical symmetry, without applying the Voronoi tessellation to the simulated data. Left: mean map obtained by averaging the 100 initial maps; right: mean map resulting from the average of the 100 maps after applying the R-L de- and reprojection. Both maps are dimensionless, have the same color scale and are shown in logarithmic scale.}
\label{fig:mapsansvor_deproj}
\end{center}
\end{figure}

\begin{figure}
\begin{center}
  \includegraphics[height=0.7\columnwidth,angle=0]{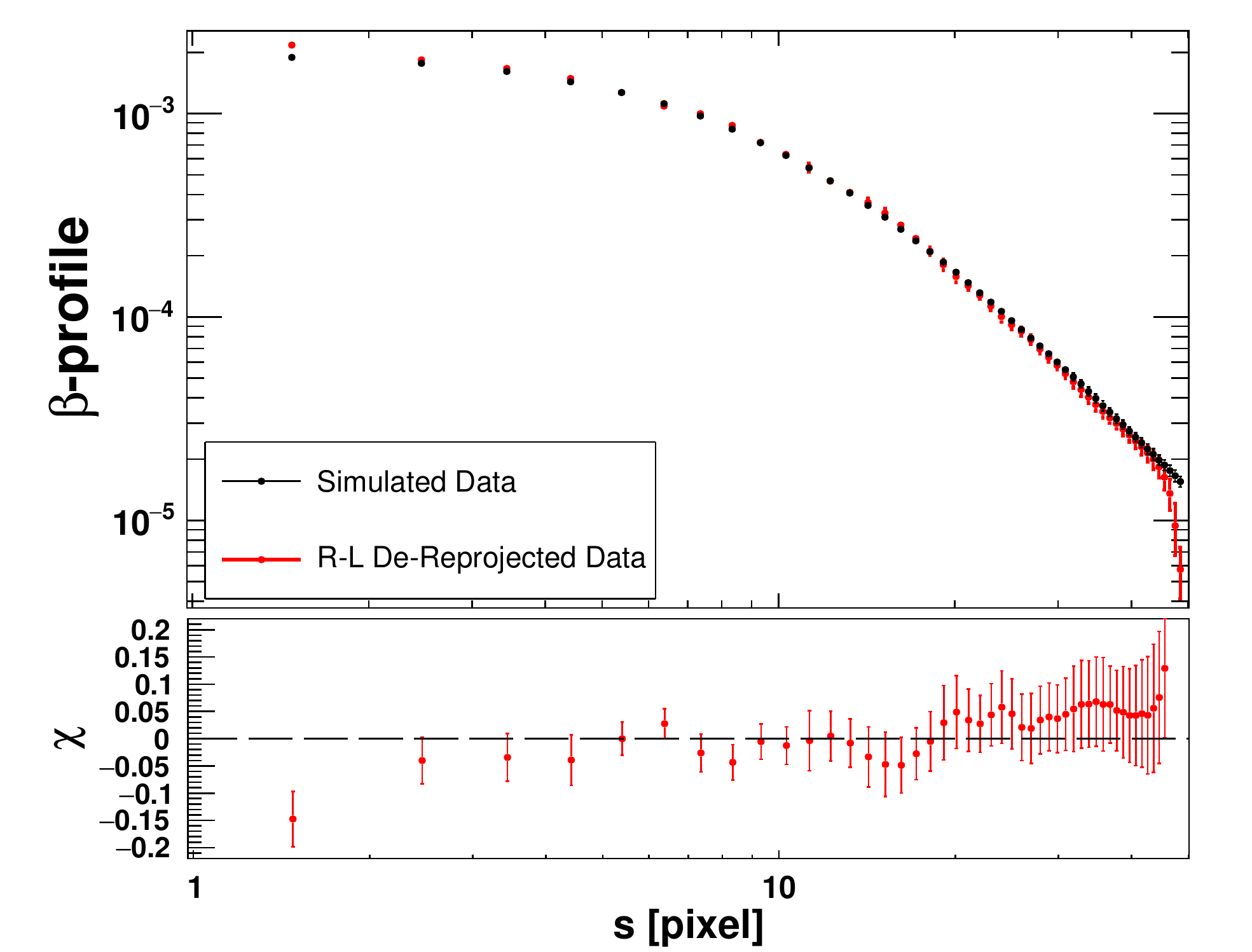}
\caption{Top: Normalized profiles illustrating the results of the de- and reprojection of the simulated $\beta$-profile map. Black: mean of the initial data maps; Red: mean of the resulting R-L deprojected and reprojected maps. These profiles correspond to the azimuthal averaged of the maps on the left and on the right of Fig.~\ref{fig:mapsansvor_deproj}, respectively. Bottom: relative residuals computed as $(f_{Data}(s)-f_{Recons}(s))/f_{Data}(s)$, with the corresponding uncertainties obtained with error propagation.}
\label{fig:sansvor_deproj}
\end{center}
\end{figure}

We first apply the R-L deprojection to the unbinned $\beta$-profile maps. The result of the R-L de- and reprojection is shown in Fig.~\ref{fig:mapsansvor_deproj}, where we show the mean of the 100 initial data maps and the mean of the 100 resulting maps, after de- and reprojection. To compare this result, in Fig.~\ref{fig:sansvor_deproj} we show profiles corresponding to an azimuthal average of the mean of the 100 initial data maps and of the mean of the 100 maps obtained after applying the R-L de- and reprojection. To produce these profiles, we derive one individual profile for each of the 100 simulated maps. From these 100 individual profiles, we compute at each radial bin the mean and the standard deviation. We then assign to each bin of the combined profile the corresponding mean and standard deviation. The other profiles shown in this study have been derived in the same way.

As we can see, the deprojected profile follows exactly the input data profile, except for the very last bin. We will return to this issue in Sect.~\ref{sec:discussion}.


\subsubsection{Deprojection of Voronoi-tessellated maps}\label{sec:betaprofvor}
\begin{figure}
\begin{center}
  \includegraphics[height=0.7\columnwidth,angle=0]{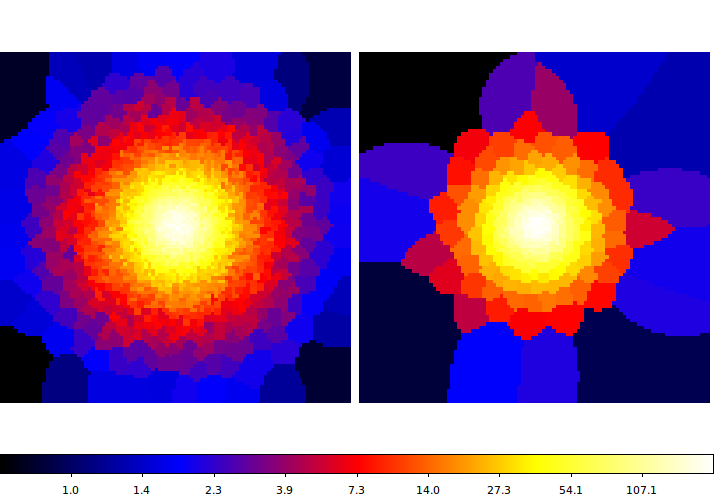}
\caption{Maps illustrating the binning with Voronoi tessellation with 30 counts/bin (left) and 300 counts/bin (right) of one noisy $\beta$-profile map (in logarithmic scale). Both maps were created with the code of \citet{eckert15_voronoi}.}
\label{fig:mapvor30_300}
\end{center}
\end{figure}

In this section, we apply Voronoi tessellation prior to R-L deprojection. To test whether the deprojection works for large smoothing at large cluster-centric distance, we require 30 counts per bin and 300 counts per bin. This illustrates the case of X-ray observations, where the signal scales with the square of the electron density, causing the outskirts of galaxy clusters to be very faint: applying Voronoi tessellation to such data leads to substantial smoothing at large radii. In Fig.~\ref{fig:mapvor30_300}, we show the initial maps after applying the Voronoi tessellation. As we can see, if the number of required counts per bin is large, the smoothing is strong at large radii. To produce these binned maps, we used the code of \citet{eckert15_voronoi}.

\begin{figure}
\begin{center}
  \includegraphics[height=0.7\columnwidth,angle=0]{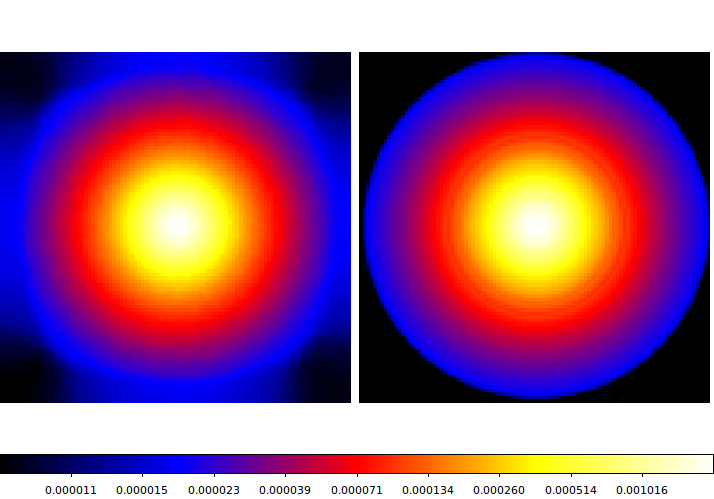}
\caption{Maps illustrating the result of the R-L deprojection algorithm on a set of 100 simulated $\beta$-profile maps (Eq.~(\ref{eq:betaprofile})) assuming spherical symmetry for a set of simulated data rebinned with a Voronoi tessellation of 30 counts/bin. Left: mean map obtained by averaging the 100 rebinned initial maps; right: mean map resulting from the average of the 100 maps after applying the R-L deprojection and reprojection. Both maps are dimensionless, have the same color scale, and are shown in logarithmic scale.}
\label{fig:mapvor30_deproj}
\end{center}
\end{figure}

\begin{figure}
\begin{center}
 \includegraphics[height=0.7\columnwidth,angle=0]{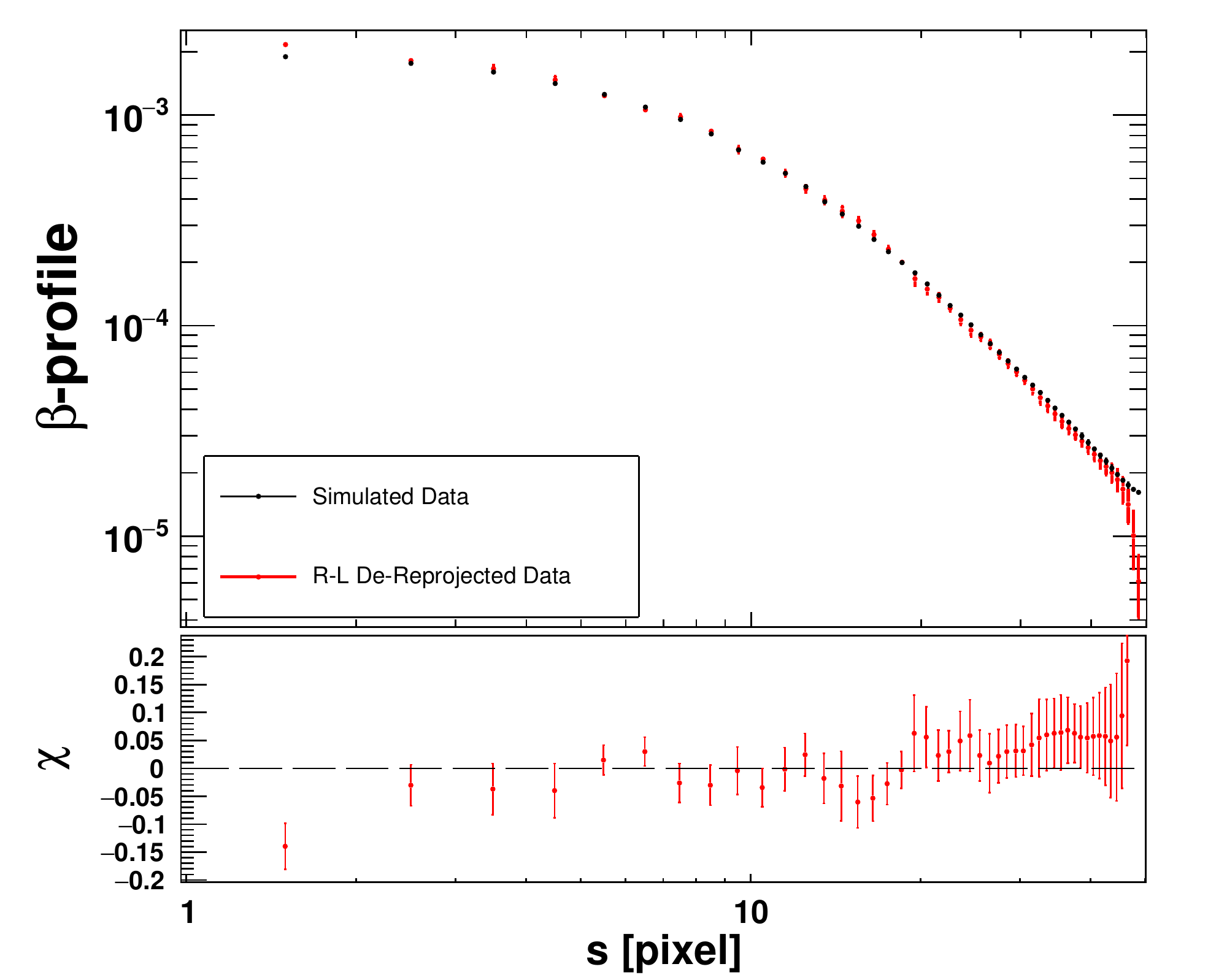}
\caption{Top: Normalized profiles illustrating the results of the de- and reprojection of the simulated $\beta$-profile map rebinned using the Voronoi tessellation with 30 counts/bin. Black: profile of the rebinned initial data maps; Red: profile obtained by de- and reprojecting the initial data maps. These profiles correspond to the azimuthal averaged of the maps on the left and on the right of Fig.~\ref{fig:mapvor30_deproj}, respectively. Bottom: relative residuals computed as in Fig.\ref{fig:sansvor_deproj}.}

\label{fig:vor30_deproj}
\end{center}
\end{figure}

\begin{figure}
\begin{center}
  \includegraphics[height=0.7\columnwidth,angle=0]{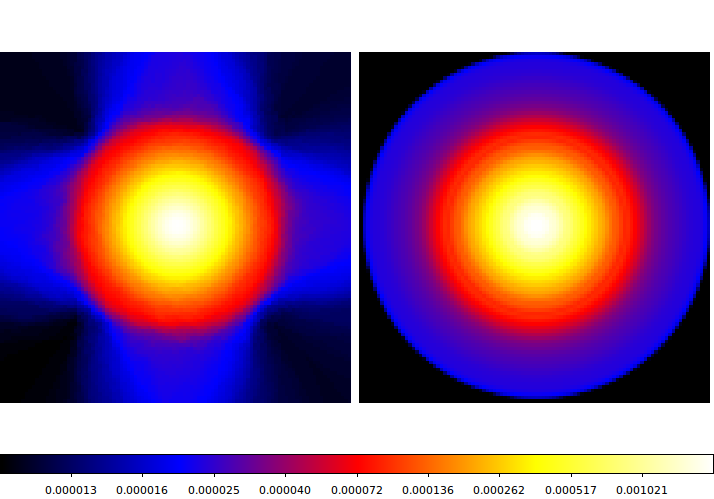}
\caption{Maps illustrating the result of the R-L deprojection algorithm on a set of 100 simulated $\beta$-profile maps (Eq.~(\ref{eq:betaprofile})) assuming spherical symmetry, using the Voronoi tessellation with 300 counts/bin applied to the simulated data. Left: mean map obtained by averaging the 100 rebinned initial maps; right: mean map resulting from the average of the 100 maps after applying the R-L deprojection and reprojection.
Both maps are dimensionless, have the same color scale, and are shown in logarithmic scale.}
\label{fig:mapvor300_deproj}
\end{center}
\end{figure}

\begin{figure}
  \includegraphics[height=0.7\columnwidth,angle=0]{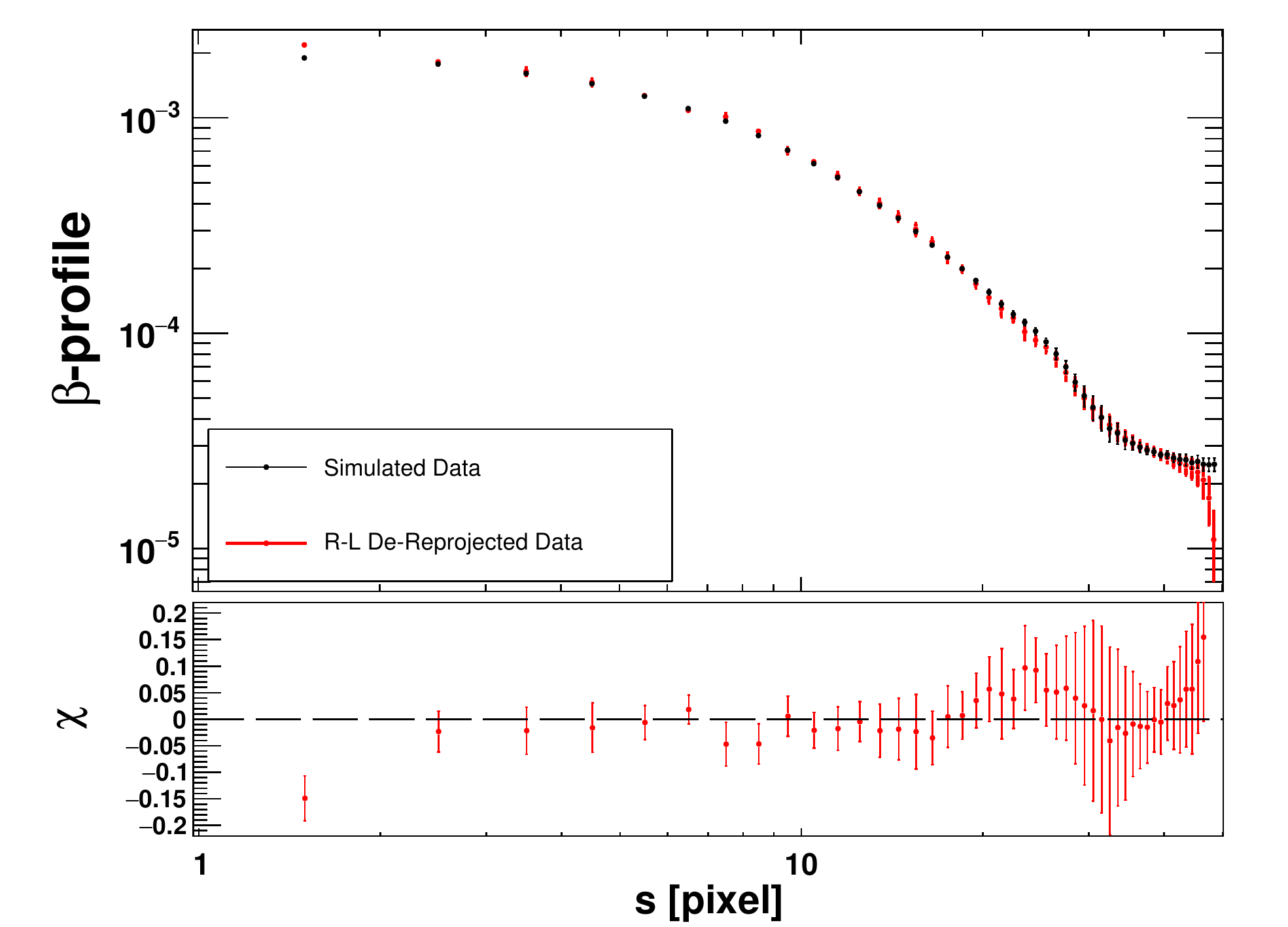}
\caption{Top: Normalized profiles illustrating the results of the de- and reprojection of the simulated $\beta$-profile map rebinned using the Voronoi tessellation with 300 counts/bin. Black: profile of the rebinned initial data maps; Red: profile obtained by de- and reprojecting the initial data maps. These profiles correspond to the azimuthal average of the maps on the left and on the right of Fig.~\ref{fig:mapvor300_deproj}, respectively. Bottom: relative residuals computed as in Fig.\ref{fig:sansvor_deproj}.}

\label{fig:vor300_deproj}
\end{figure}

We now proceed to test the deprojection method on these maps. We start by simulating 100 noisy $\beta$-profile maps, which we bin using Voronoi tessellation \citep{eckert15_voronoi}. Then, we deproject each of these 100 binned maps and reproject them. For a tessellation with 30 counts/bin, the map resulting from the mean of the 100 R-L deprojected and reprojected maps is shown in the right panel of Fig.~\ref{fig:mapvor30_deproj}; while the mean of the initial 100 maps is shown in the left panel. To help the interpretation of the results, we azimuthally average these maps to produce the profiles shown in Fig.~\ref{fig:vor30_deproj}. In Figs.~\ref{fig:mapvor300_deproj} and \ref{fig:vor300_deproj}, we show the results obtained for the data binned requiring 300 counts/bin. The left panel of Fig.~\ref{fig:mapvor300_deproj} shows an interesting pattern which is due to the large smoothing by the Voronoi binning. Indeed, it represents the mean of the 100 noisy maps binned in a similar way as the map shown in the right panel of Fig.~\ref{fig:mapvor30_300}.

We can see that the deprojection works quite well over the entire profile up to the very last bin, which again decreases too fast compared to the initial data set. 

\section{Reconstruction of the 2D gravitational potential} \label{sec:recons_method}

Here, we outline the method to reconstruct the gravitational potential of galaxy clusters from the cluster observables. We shall focus on the cluster gravitational-potential reconstruction using the X-ray emission of the galaxy clusters \citep[see ][ and for an application to the cluster Abell 1689, \citet{tchernin15}]{konrad13} and using the SZ signal \citep{majer13}. The reconstruction of the gravitational potential from the velocity dispersion of the galaxies in the cluster \citep{sarli14}, whose more complicated physics requires a deprojection kernel that differs from the one of Eq.~(\ref{eq:sphk}), will be considered in a separate and dedicated study.
\\\\To keep the covariance matrices easily manageable, they are limited to initial data sets of approximately 100x100 pixels. Thus, we rebin the grid used in this study to satisfy this condition.
\subsection{Method for reconstructing the cluster gravitational potentials using X-ray data}
The ICM is filled with hot plasma making clusters bright extended X-rays sources due to thermal bremsstrahlung and line emission. 

Assuming hydrostatic equilibrium, information on the cluster gravitational potential can be obtained from the physical properties of the ICM. In hydrostatic equilibrium, the ICM gas pressure $P$ and the gravitational potential $\Phi$ are related by
\begin{equation}\label{eq:eqHydro}
  \nabla P=-\rho \nabla\Phi\;,
\end{equation}
where $\rho$ is the gas density. All these quantities have a radial dependence, not shown explicitly here for improved readibility.

We further assume that the plasma follows the polytropic relation
\begin{equation}\label{eq:eqPoly}
  \frac{P}{P_0}=\left(\frac{\rho}{\rho_0}\right)^{\gamma}\;,
\end{equation}
where the suffix $0$ corresponds to the value of the pressure and of the gas density at an arbitrary fiducial radius $r_0$, and $\gamma$ is the polytropic exponent. Finally, we assume that the gas is ideal,
\begin{equation}\label{eq:idealGas}
  P=\frac{\rho}{\bar{m}} k_\mathrm{B}T\;,
\end{equation}
where $T$ is the gas temperature, $\bar{m}$ is the mean mass of a gas particle, and $k_\mathrm{B}$ is Boltzmann's constant.

The bolometric bremsstrahlung emissivity, $j_x$, depends on temperature and density as
\begin{equation}\label{eq:Brem}
  j_x\propto T^{1/2}\rho^2\;.
\end{equation}
Assuming that the thermal bremsstrahlung dominates the X-ray emission, the X-ray emissivity can  be related to the 3D gravitational potential $\Phi$, as derived in \citet{konrad13},
\begin{equation}\label{eq:phi_x}
  \Phi\propto j_x^{1/\eta}\;,\quad\eta=\frac{3+\gamma}{2(\gamma-1)}\;.
\end{equation}

This method returns an estimate of the 3D Newtonian gravitational potential $\Phi$ \citep[using Eq.(6) of ][]{konrad13}. We point out that we do not need to extract the density and temperature profiles of a cluster to reconstruct its potential, but only the emissivity profile, which combines both quantities. To compare this reconstructed gravitational potential with the observed lensing potential, we reproject it according to
\begin{equation}\label{eq:Phi_newton}
  \Psi(s)\propto \int \Phi(r)dz\;,
\end{equation}
where $s$ is the projected radius, $z$ the line-of-sight and $r$ the 3D radius.
\\We note that we only have access to the bremsstrahlung emissivity projected along the line-of-sight, that is,\ the surface brightness \textit{SB}$_x$($s$)=$\int dz j_x(r)$. Therefore, before applying Eq.~(\ref{eq:phi_x}) we first use the R-L method to deproject the surface brightness \textit{SB}$_x$($s$) and recover the emissivity $j_x$($r$). 

\subsection{Method for reconstructing the gravitational cluster potentials using SZ observations}
When cosmic microwave background (CMB) photons pass through galaxy clusters, they can be upscattered by inverse Compton scattering off the thermal electrons in the intracluster plasma. Their energy gain leaves an imprint on the photon energy distribution: the energy spectrum of a small fraction of the photons is blue-shifted with respect to the CMB spectrum. This is called the (thermal) Sunyaev-Zel'dovich (SZ) effect, which is observable at millimeter wavelengths. The signature left on the CMB spectrum is proportional to the electron pressure in the ICM gas integrated along the line-of-sight, the so-called Compton-$y$ parameter
\begin{equation}
  y(s)=\frac{k_\mathrm{B}}{m_ec^2}\sigma_T\int dz T(r)\rho(r)\;,
\end{equation}
where $\sigma_T$ is the Thomson cross-section, $m_e$ the electron mass and $T(r)\rho(r)$ the electron pressure.

Similar to the X-ray emission, the information provided by the SZ signal can be used to estimate the 3D gravitational potential. The assumption on the plasma properties are the same as in Eqs.~(\ref{eq:eqHydro})-(\ref{eq:idealGas}).

Once the Compton-$y$ parameter has been deprojected to recover the electronic gas pressure $P$, the 3D gravitational potential can be reconstructed from \citep{majer13}
\begin{equation}\label{eq:phi_sz}
  \Phi\propto P^{1/\eta}\;,\quad\eta=\frac{\gamma}{(\gamma-1)}\;.
\end{equation}

This estimate of the Newtonian gravitational potential is then reprojected following Eq.~(\ref{eq:Phi_newton}) to be compared to the observed lensing potential.

\section{Reconstruction of the projected gravitational potential of the simulated g1-cluster}\label{sec:recons_g1}

In this section, we derive the projected gravitational potential of a simulated cluster at redshift 0.297 for which we have the X-ray, SZ signal, and the true projected potential \citep[the g1-cluster,][]{meneghetti10}. The simulated maps cover a field of view of 4 Mpc/h x 4 Mpc/h and the assumed cosmology is $\Lambda$CDM with $\Omega_m=0.3$, $\Omega_{\Lambda}$=0.7 and H$_{0}$=70 km/(s Mpc). This simulation is actually the result of a re-simulation of the numerical hydrodynamical simulations presented by \citet[][]{saro06}, itself also extracted from a dark-matter-only simulation from \citet{yoshida01}. This simulation, re-simulated with the N-body/SPH-code Gadget 2 \citep{springel05} includes both dark matter and baryonic physics \citep[e.g., gas cooling, star formation, chemical enrichment, see][for more details]{meneghetti10}.

The initial images have a dimension of 512x512 pixels, each pixel having a size of 11.16~kpc.
We bin these maps by 8 pixels to avoid potential problems in memory allocation for the computation of the covariance matrix (Sect.~\ref{sec:join}).  We will describe how we derived the 100 maps from the X-ray and SZ mock data below.

\subsection{Simulated X-ray data}
The X-ray emission is provided in erg/s for the energy band 0.13-13~keV. We simulate \textit{XMM-Newton} mock observations using the PIMMS interface\footnote{\url{https://heasarc.gsfc.nasa.gov/cgi-bin/Tools/w3pimms/w3pimms.pl}}(A Mission Count Rate Simulator). For this purpose, we assume that the thermal emission of this cluster follows a thin plasma model \textit{APEC} \citep{smith01} with a metal abundance of 0.2 the Solar value, and a temperature of 5 keV (motivated by the averaged emitted power which is of the order of $10^{44}$ erg/s). This allows us to convert the flux in erg/(s cm$^2$) to the number of counts that the \textit{XMM-Newton} satellite would detect in the energy band 0.5-2keV assuming an exposure time of 100ks. The number of counts ($n$) in each pixel is expected to follow a Poisson distribution, with a standard deviation of $\sqrt{n}$. We assume that the sky background is at the level estimated in \citet{tchernin16a} in the specific case of Abell 2142, that is, 5\% of the sky background components  (composed of the cosmic X-ray background (CXB), the Galactic halo, and the local hot bubble). For the non-X-ray background (NXB) we use the model estimated for A2142 in \citet{tchernin16a}, that we correct for the exposure time of the simulation. For consistency, we place here the g1-cluster at the redshift of Abell 2142. 

We generate 100 maps resembling these mock data assuming that the value at each pixel is drawn from a Poisson distribution that accounts for both the statistical and the systematic uncertainties.

\subsection{Simulated SZ data}
We produce NIKA2-like mock observations of the g1-cluster. NIKA2 is a ground-based camera\footnote{\url{http://ipag.osug.fr/nika2/Instrument.html}} for microwave and submillimeter observations at 150 and 260 GHz \citep{calvo16} with an angular resolution of 18.5 arcsec full width at half maximum (FWHM) at 150 GHz (the effect of the PSF corresponds to the size of a pixel of the rebinned map and is therefore negligible). This choice is motivated by the small angular size of the g1-cluster on the sky, which would make it appear point-like for \textit{Planck}-like observations at 5-10 arcmin FWHM angular resolution.  For the simulation, we convolve the SZ signal map of the g1-cluster with a Gaussian point-spread function of 18.5 arcsec  FWHM. The variance of the white Gaussian noise is inversely proportional to the exposure time and equal to $1.6 \cdot 10^{-9}$ per beam for one hour of observation. In the present analysis, we generate 100 simulated maps of the  SZ emission assuming an exposure time of $4$ hr. We note that the 6.5 arcmin field of view of NIKA2 corresponds to roughly $\ell =
3000$ in the CMB power spectrum.  At those scales, the noise due to CMB anisotropies (with a variance of the order of 10$^{-14}$ in Compton
parameter) is negligible compared to the instrumental noise.

\subsection{Deprojection of the simulated X-ray and SZ signal}

\begin{figure}
\begin{center}
  \includegraphics[height=0.5\columnwidth,angle=0]{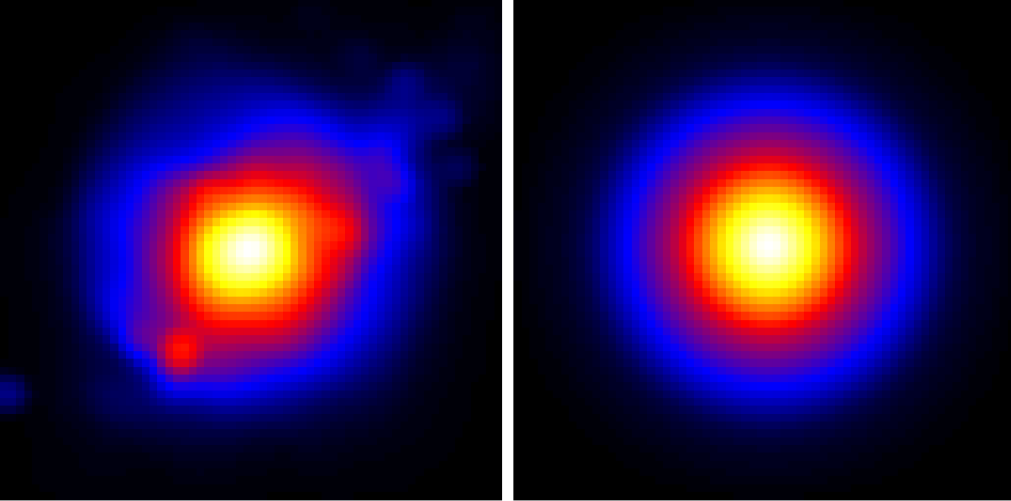}
\caption{Maps illustrating the result of the R-L deprojection algorithm on a set of 100 simulated maps of the X-ray emission of the g1-cluster assuming spherical symmetry. Left: mean map obtained by averaging the 100 simulated X-ray maps of the g1-cluster; right: mean map resulting from the average of the 100 maps obtained after applying the R-L de- and reprojection.
Both maps are dimensionless, have the same color scale (ranging from 2e-5 to 8.30e-3, from dark blue to light yellow), and are shown in logarithmic scale.}
\label{fig:G1MapXR_deproj}
\end{center}
\end{figure}

\begin{figure}
\begin{center}
  \includegraphics[height=0.7\columnwidth,angle=0]{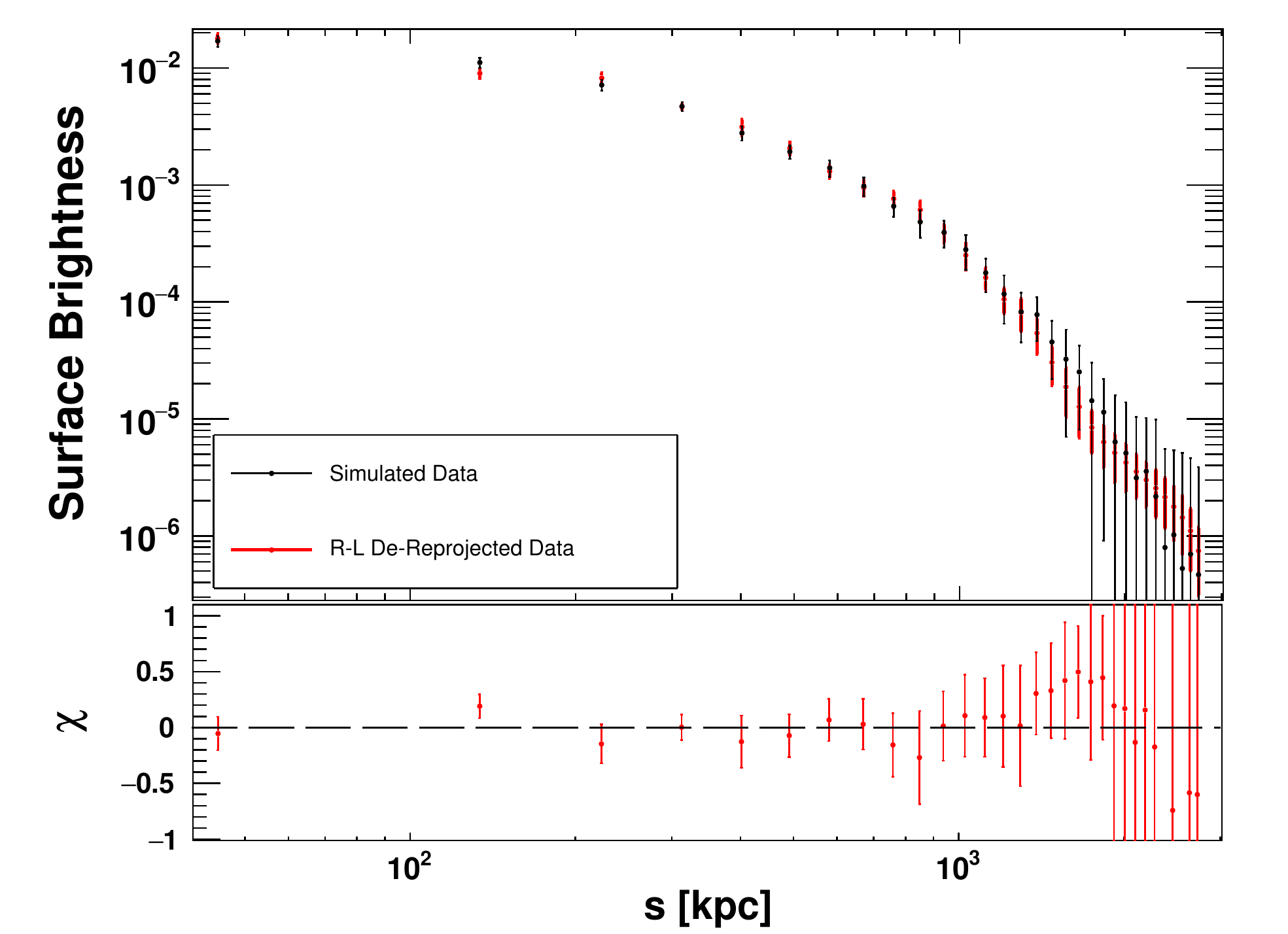}
\caption{Top: Normalized profiles illustrating the results of the de- and reprojection of the X-ray emission of the g1-cluster. Red: profile of the initial data maps; blue: profile obtained after de- and reprojection. These profiles correspond to the azimuthal average of the maps on the left and on the right of Fig.~\ref{fig:G1MapXR_deproj}, respectively. Bottom: relative residuals computed as in Fig.\ref{fig:sansvor_deproj}.}
\label{fig:G1XR_deproj}
\end{center}
\end{figure}

Assuming spherical symmetry, the result of the R-L de- and reprojection of 100 simulated X-ray maps is shown in Fig.~\ref{fig:G1MapXR_deproj}. There, we show the average of these 100 maps before (left panel) and after (right panel) the de- and reprojection. In Fig.~\ref{fig:G1XR_deproj}, we show the azimuthal profiles obtained from the average maps shown in Fig.~\ref{fig:G1MapXR_deproj}. 

\begin{figure}
\begin{center}
  \includegraphics[height=0.5\columnwidth,angle=0]{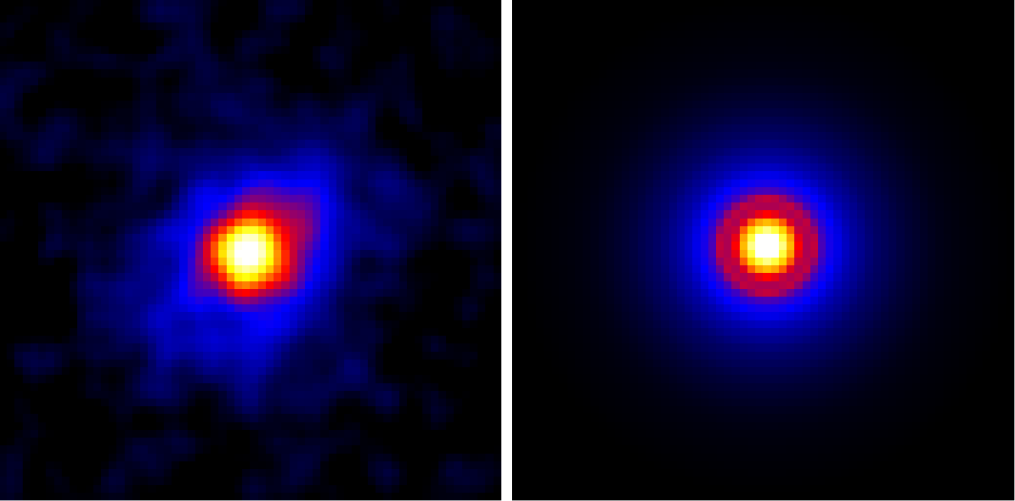}
\caption{Maps illustrating the result of the R-L deprojection algorithm on a set of 100 simulated SZ maps of the g1-cluster assuming spherical symmetry. Left: mean map obtained by averaging the 100 simulated SZ maps of the g1-cluster; right: mean map resulting from the average of the 100 maps obtained after applying the R-L de- and reprojection. Both maps are dimensionless and have the same color scale (ranging from 6.0e-4 to 5.45e-3, from dark blue to light yellow).}
\label{fig:G1MapSZ_deproj}
\end{center}
\end{figure}

\begin{figure}
\begin{center}
  \includegraphics[height=0.7\columnwidth,angle=0]{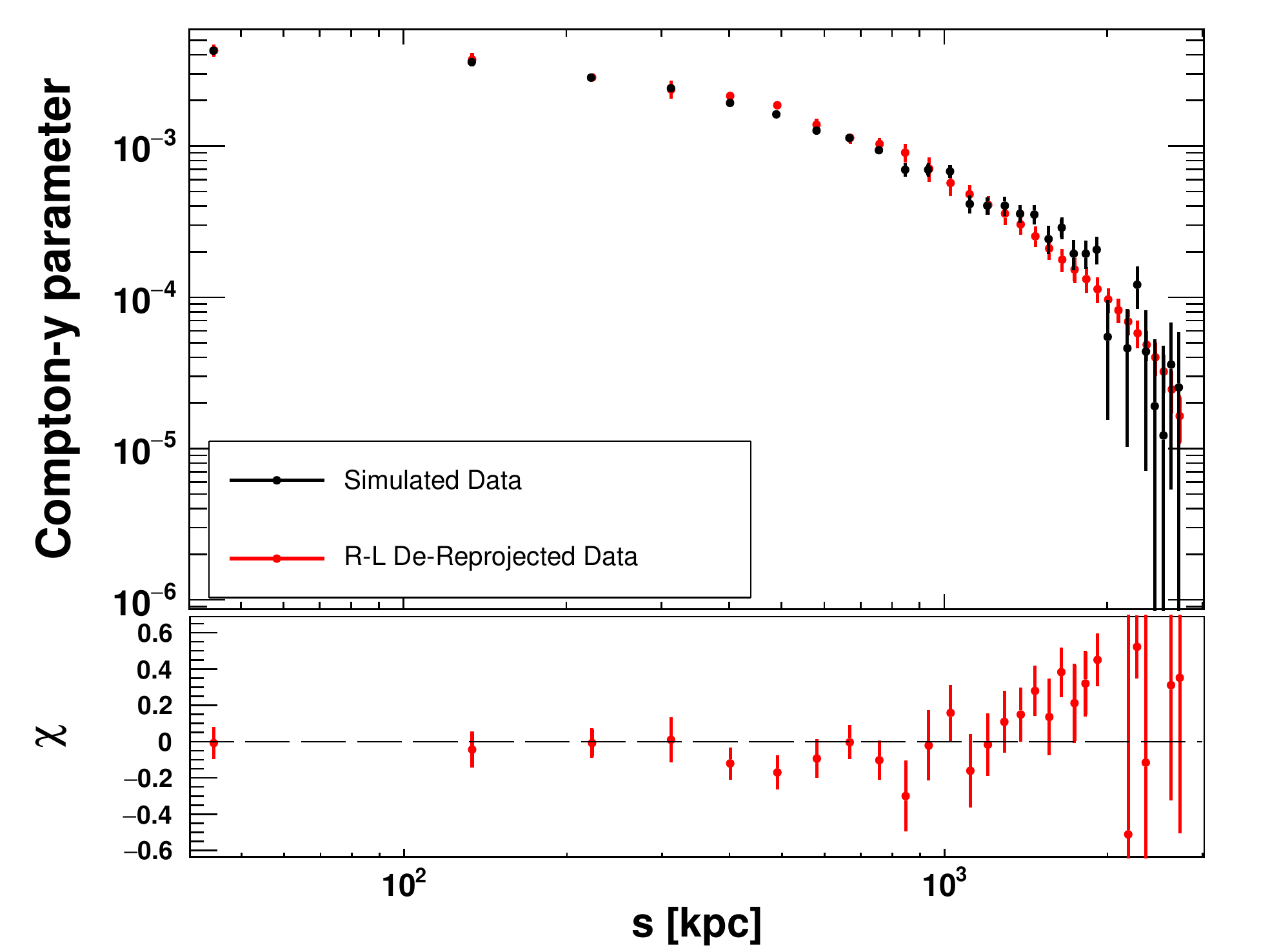}
\caption{Top: Normalized profiles illustrating the results of the de- and reprojection of the SZ emission of the g1-cluster. Black: profile of the initial data maps; Red: profile obtained after de- and reprojection. These profiles correspond to the azimuthal averaged of the maps on the left and on the right of Fig.~\ref{fig:G1MapSZ_deproj}, respectively. Bottom: relative residuals computed as in Fig.\ref{fig:sansvor_deproj}.}
\label{fig:G1SZ_deproj}
\end{center}
\end{figure}

We then apply the same procedure to the simulated SZ signal. The resulting averaged maps are shown in Fig.~\ref{fig:G1MapSZ_deproj}, before (left panel) and after (right panel) the de- and reprojection, while the azimuthally averaged profiles of these maps are shown in Fig.~\ref{fig:G1SZ_deproj}.

We note that the de- and reprojected quantities are normalized during the deprojection procedure and that therefore only the overall shape of the profiles in Figs.~\ref{fig:G1MapXR_deproj} and~\ref{fig:G1MapSZ_deproj} contains a physical meaning.

\subsection{Reconstruction of the projected gravitational potential from the X-ray and SZ signals of the simulated g1-cluster}
\begin{figure}
\begin{center}
  \includegraphics[height=0.7\columnwidth,angle=0]{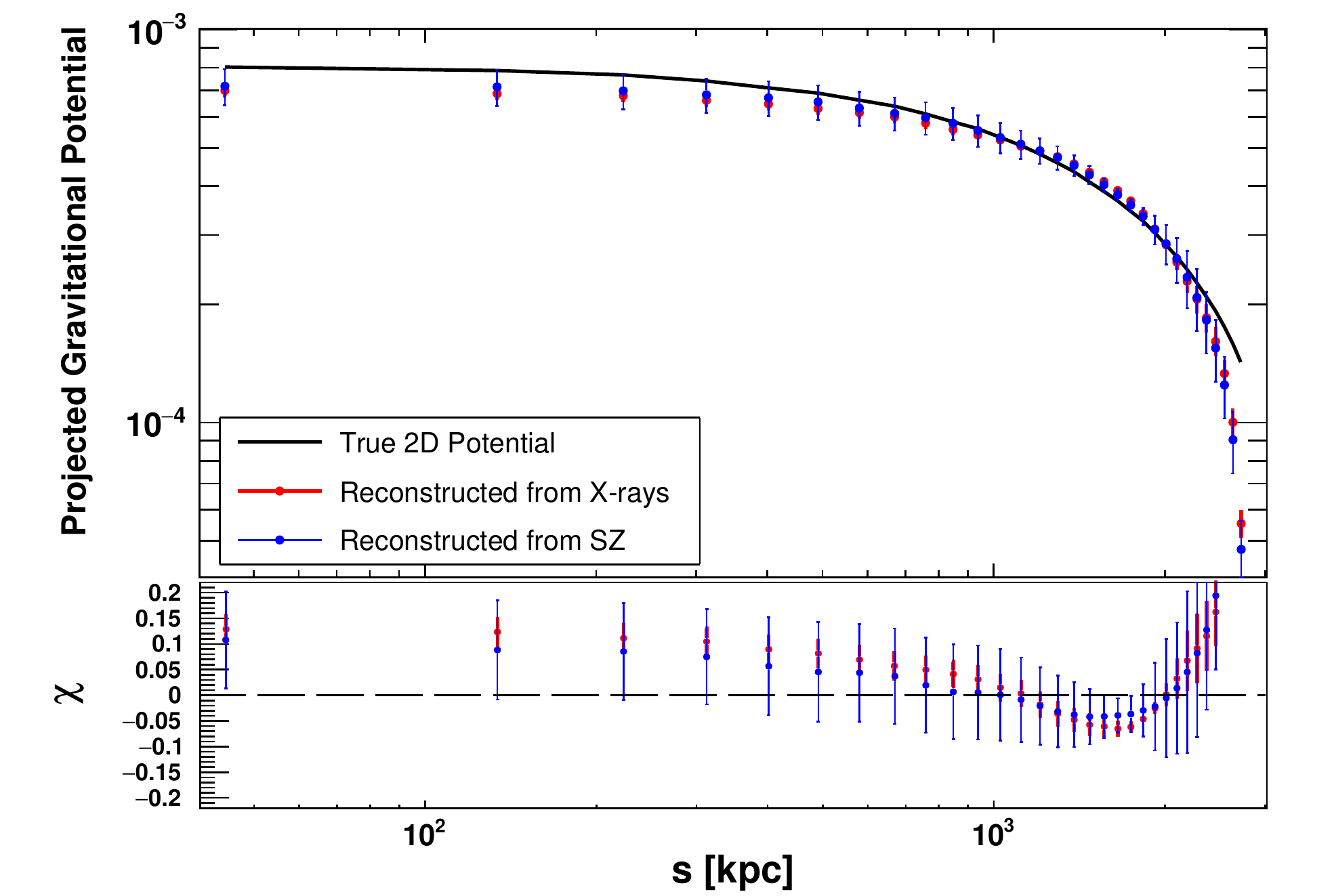}
\caption{Top: Normalized azimuthally averaged projected gravitational potential reconstructed from the X-ray (Eq.~(\ref{eq:phi_x})) and from the SZ signal (Eq.~(\ref{eq:phi_sz})) of the g1-cluster, compared to the true projected potential. Blue: projected potential reconstructed from the SZ signal; Red: projected potential reconstructed from the X-ray signal; black: true projected potential. Bottom: relative residuals computed as $(f_{true}(s)-f_{Recons}(s))/f_{true}(s)$, with the corresponding uncertainties obtained with error propagation and no uncertainties on $f_{true}(s)$.}
\label{fig:G1_lensing} 
\end{center}
\end{figure}

We now reconstruct the projected gravitational potential from the X-ray and SZ signal of the g1 cluster, using Eqs.~(\ref{eq:phi_x}) and (\ref{eq:phi_sz}), respectively. The 2D potentials recovered this way for a polytropic index of 1.22 \citep{meneghetti10} are shown in Fig.~\ref{fig:G1_lensing}. The reconstructed profiles are compared to the true projected potential. To simplify this comparison, all projected potentials have been set to zero at the virial radius \citep[$R_{vir}\sim$ 2600kpc,][]{meneghetti10}. We note that a shift by a constant will not affect our joint reconstruction, as it aims at reproducing the lensing observables, which are combinations of second derivatives of the projected potential (see Sect.~\ref{sec:cov_computation}).

The residuals of the reconstruction can be seen in the bottom panel of Fig.~\ref{fig:G1_lensing}. While the potential recovered from the SZ signal agrees with the true potential, the residuals between the true potential and the potential reconstructed from the X-ray data show larger variations.
This is largely due to the sizes of the error bars which are different in the potentials reconstructed from the X-ray and from the SZ measurements. Interestingly, we can see the same wiggly feature in the potential reconstructed from SZ and X-rays data. This could be a hint that the hydrostatic equilibrium may not be valid in some regions of the cluster and therefore, such a comparison can contain valuable information about the physical state of the cluster.  We will return to this point in the discussion.

\section{Reconstruction of the projected gravitational potential of Abell 2142} \label{sec:recons_A2142}
Abell 2142 is a massive cluster \citep[$M_{200}\sim 1.3\cdot 10^{15}M_{\sun}$, ][]{munari14} located at a redshift of 0.09. There is evidence revealing it as a dynamically active cluster: it is accreting substructure \citep{eckert1a2142}, and sloshing activity has been observed to be ongoing in the central region out to 1~Mpc \citep{rossetti13}. Furthermore, the study of the galaxy distribution by \citet{owers11} indicates the presence of minor mergers in the cluster. Nevertheless, there  are no hints for this cluster to be out of hydrostatic equilibrium in its outskirts \citep{tchernin16a}. Therefore, this cluster may be a good candidate to apply the potential reconstruction from X-ray and SZ measurements. We will discuss the validity of the assumptions for this cluster in more details in Sect.~\ref{sec:ass_disc}.

\subsection{X-ray observations of Abell 2142 with \textit{XMM-Newton}}
\begin{figure}
\begin{center}
  \includegraphics[height=0.45\columnwidth,angle=0]{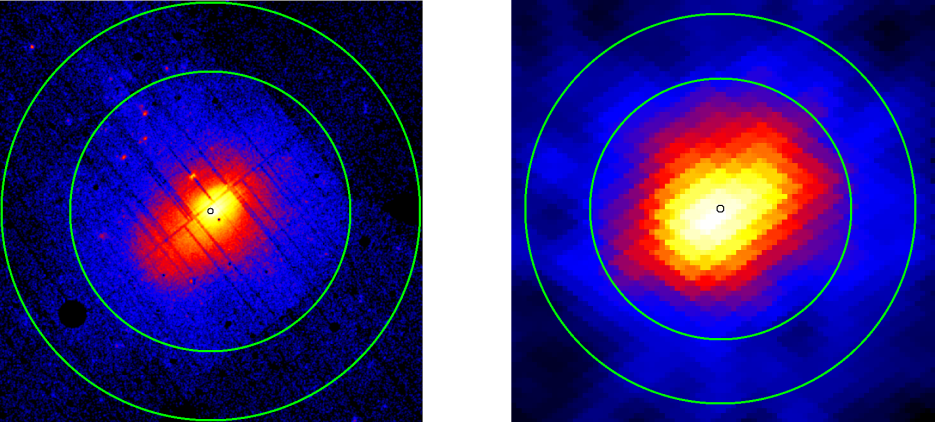}
\caption{Regions of the observations of Abell 2142 used in this analysis. Left: \textit{XMM-Newton} X-ray data in counts (ranging from 0, in black; to 120 counts, in white), taken within the X-COP program \citep{eckert16_xcop} (in logarithmic scale); right: \textit{Planck} SZ data in units of the Compton-$y$ parameter (ranging from 8.5e-7, in black; to 6.6e-5 counts, in white), using the method of \citet{hurier13}. The black circle represents the position of the image center used in our analysis, while the two green circles outline $R_{500}$ and $R_{200}$.}
\label{fig:A2142_data}
\end{center}
\end{figure}

We use the data collected within the X-COP project \citep[PI: D. Eckert,][]{eckert16_xcop}. 
Owing to the pointing strategy of the X-COP program, the data extend beyond $R_{200}$. We first center the map on the cluster center (at the position Ra: 239.5858 deg and Dec: 27.2270 deg) and cut the image to 1000x1000 pixels (each pixel has a size of 4.25~kpc), limiting our field-of-view to $\sim$4250~kpc x $\sim$4250~kpc. This allows us to analyze the cluster up to a radius of $\sim$ 2125~kpc \citep[which is of the order of $R_{200}$ for this cluster, see e.g., Table 4 in][]{tchernin16a}. The initial data map used here is shown in the left panel of Fig.~\ref{fig:A2142_data}. The black circle represents the position of the center of the image that has been used in the analysis, while the two green circles are estimates of $R_{500}$ ($\sim$0.66$R_{200}\approx$  1402~kpc) and of $R_{200}$ (for the smaller and the larger circle, respectively).  The systematic \text{errors} for this observation amount to 5\% of the sky background components \citep[see Appendix B of][]{tchernin16a}.

\begin{figure}
\begin{center}
  \includegraphics[height=0.5\columnwidth,angle=0]{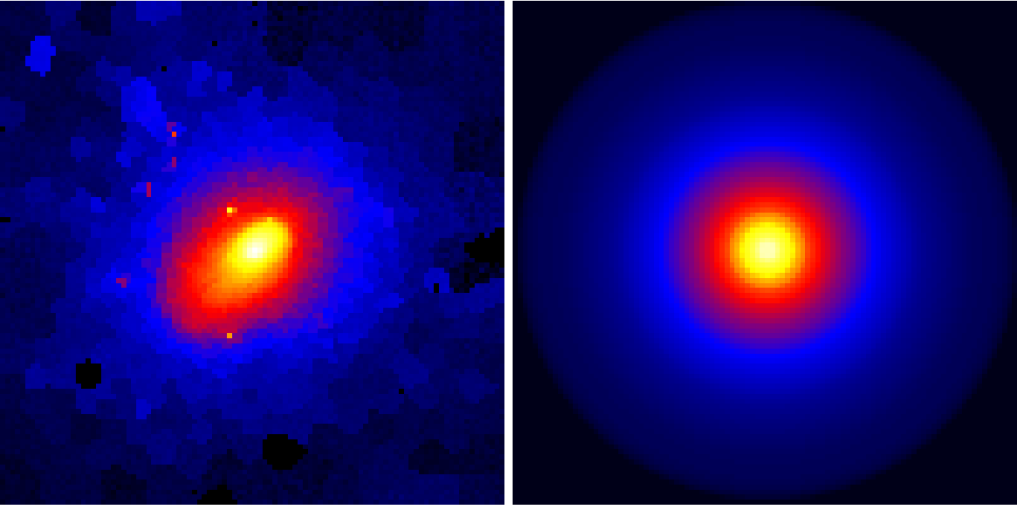}
\caption{Maps illustrating the result of the R-L deprojection algorithm on a set of 100 simulated X-ray maps of the Abell 2142 emission (simulated from the observations of XMM-Newton) assuming spherical symmetry. Left: mean map obtained by averaging the 100 simulated X-ray maps of Abell 2142; right: mean map resulting from the average of the 100 maps obtained after applying the R-L de- and reprojection. Both maps are dimensionless, have the same color scale (ranging from 2e-5 to 8.05e-3, from dark blue to light yellow) and are shown in logarithmic scale.}
\label{fig:MapXR_A2142_deproj}
\end{center}
\end{figure}

We rebin this map by averaging over 10x10 pixel blocks to satisfy the size limit set for computing the covariance matrix. Thus, each pixel of the map we are using has a $\sim$42.5~kpc x $\sim$42.5~kpc size and contains the photon counts of 100 pixels in the original data. We then apply the Voronoi code by \citet{eckert15_voronoi} to ensure 200 cts/bin. To simulate the 100 maps from this observation, we assume that the photon counts in each pixel follow a Gaussian distribution whose mean is provided by the Voronoi code of \citet{eckert15_voronoi} and whose standard deviation accounts for both the statistical and the systematic uncertainties. The mean map resulting from these 100 maps is represented on the left panel of Fig.~\ref{fig:MapXR_A2142_deproj}.

\begin{figure}
\begin{center}  
\includegraphics[height=0.7\columnwidth,angle=0,angle=0]{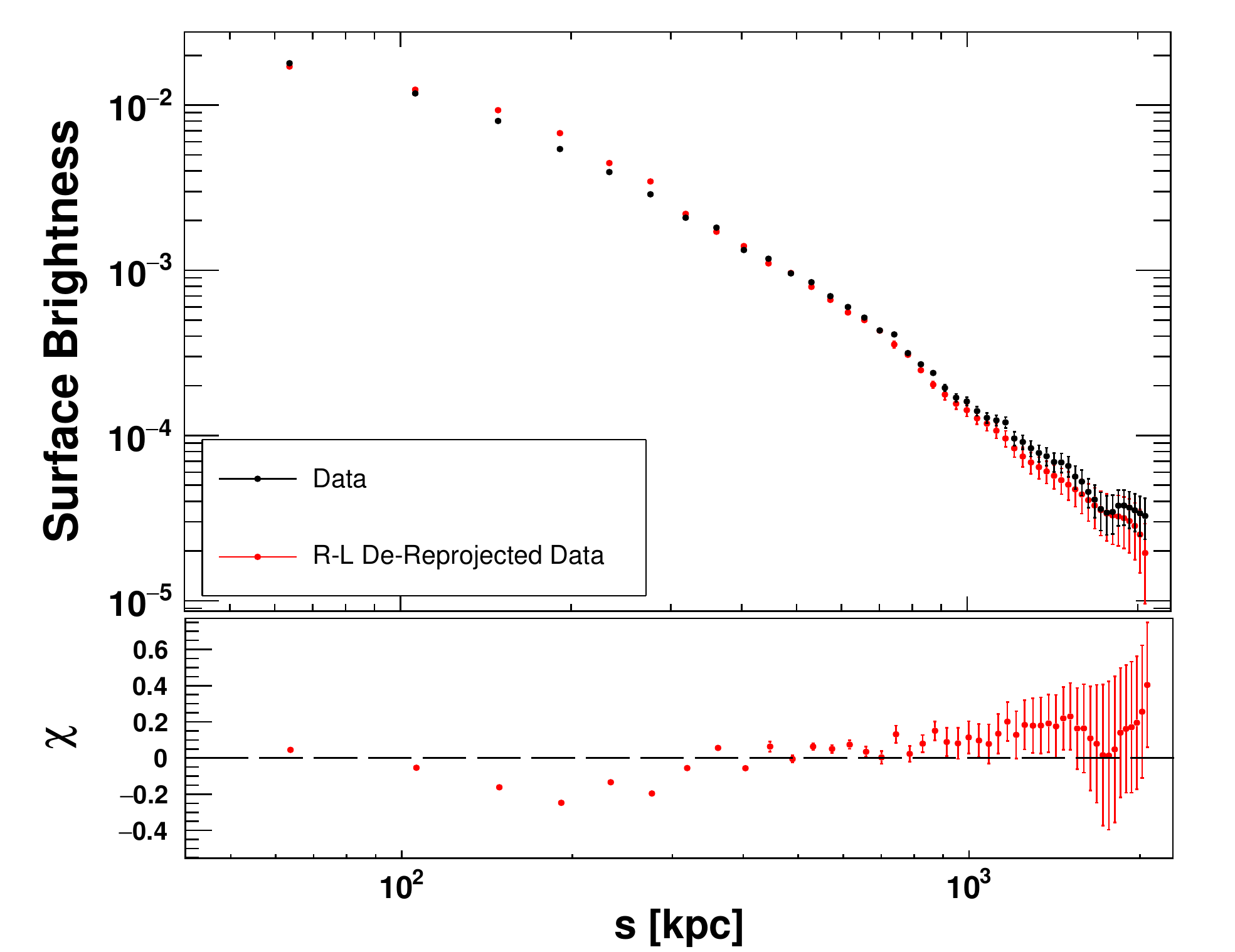}
\caption{Top: Normalized profiles illustrating the results of the de- and reprojection of the X-ray emission of Abell 2142. Black: profile of the initial data maps; Red: profile obtained by de- and reprojecting the data maps. These profiles correspond to the azimuthal average of the maps on the left and on the right of Fig.~\ref{fig:MapXR_A2142_deproj}, respectively. Bottom: relative residuals computed as in Fig.\ref{fig:sansvor_deproj}.}

\label{fig:XR_A2142_deproj}

\end{center}
\end{figure}

We then de- and reproject each of these 100 maps. 
The result of this operation on the X-ray signal from Abell 2142 is shown in the right panel of Fig.~\ref{fig:MapXR_A2142_deproj}, for the averaged map and in Fig.~\ref{fig:XR_A2142_deproj}, for the azimuthally averaged profiles.
We note that the PSF of \textit{XMM-Newton} can be characterized by a FWHM\footnote{\url{http://heasarc.nasa.gov/docs/xmm/uhb/onaxisxraypsf.html}, table 2} of $6.6$ arcsec ($\sim$11.22 kpc for this cluster) for pn and with a smaller FWHM for the two other EPIC instruments (with 6.0 and 4.5 arcsec, for MOS-1 and MOS-2 respectively). The effect of the PSF is thus smaller than the size of a pixel of the rebinned grid and is negligible.

We point out that the low resolution of the rebinned grid should not be an issue in the joint reconstruction because the joint reconstruction is based on the combination with weak lensing data, whose resolution is of the order of 100kpc and with strong lensing data, which supply high-resolution information at the cluster center. We will return to this point in the discussion.
 
\subsection{SZ observations of Abell 2142 with \textit{Planck}}
We use the millimeter observations taken by the \textit{Planck} satellite \citep{planck13}.
This map has been extracted at the frequencies 70 to 857 GHz using the MILCA astrophysical component separation method \citep{hurier13}.
It has also been used in the analysis perfomed in \citet{tchernin16a}, and we refer the reader interested in the details of the extraction of this SZ map to Sect. 3.4 of that paper.
We note that the noise in MILCA SZ maps is Gaussian, correlated, and inhomogeneous.

The map initially had a side length of $20\,R_{500}$, but we cut it to approximately the dimension of the X-ray map (i.e., to a map of 166x166 pixels, with each pixel having 27.48 kpc size and we rebinned them by 2 to avoid memory allocation issues in the covariance matrix computation). 
The initial map used in this analysis is shown in the left panel of Fig.~\ref{fig:A2142_data}: the black circle represents the image center used in our analysis, while the two green circles outline our estimates of $R_{500}$ and $R_{200}$. As we can see, the SZ map extends to slightly larger distances from the center than the X-ray map ($\sim$2125 kpc for the X-ray field-of-view compared to $\sim$2280 kpc for the SZ field-of-view).

\begin{figure}
\begin{center}
  \includegraphics[height=0.5\columnwidth,angle=0]{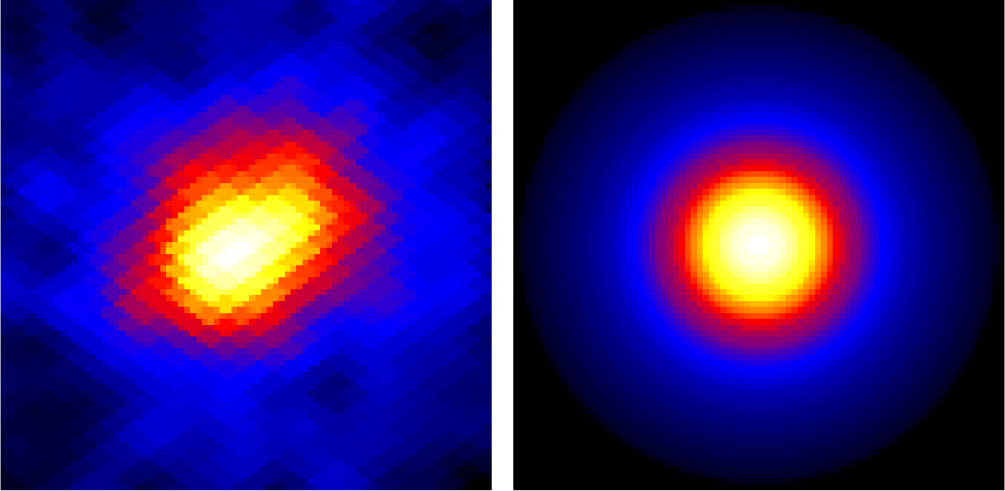}
\caption{Maps illustrating the result of the R-L deprojection algorithm on a set of 100 simulated maps drawn from the SZ data of Abell 2142 collected by the \textit{Planck} satellite assuming spherical symmetry. Left: mean map obtained by averaging the 100 simulated maps of the X-ray emission of Abell 2142; right: mean map resulting from the average of the 100 maps obtained after applying the R-L de- and reprojection. Both maps are dimensionless and have the same color scale (ranging from 1.06e-4 to 7.05e-4, from dark blue to light yellow).}\label{fig:MapSZ_A2142_deproj}
\end{center}
\end{figure}

\begin{figure}
\begin{center}
  \includegraphics[height=0.7\columnwidth,angle=0]{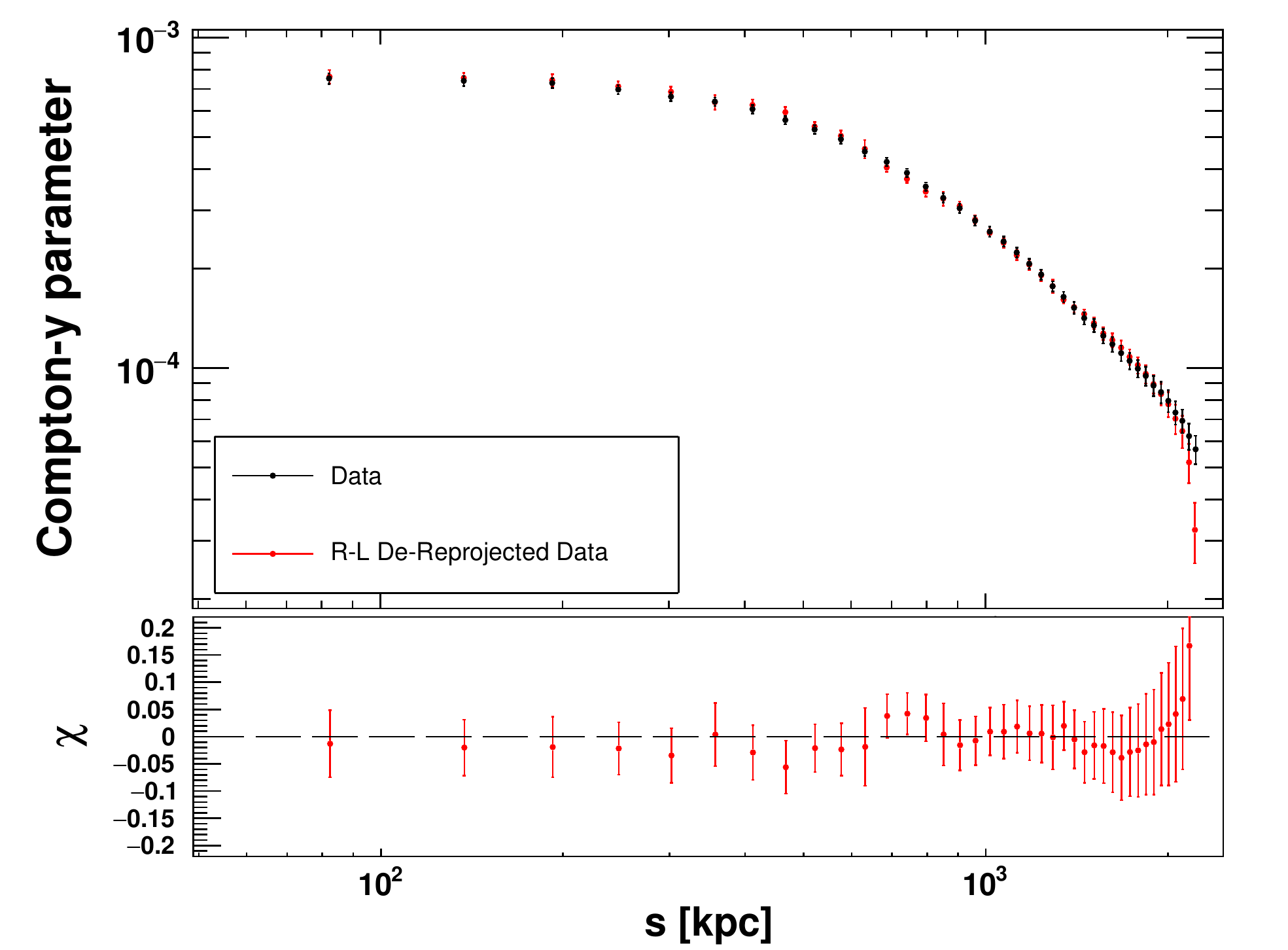}
\caption{Top: Normalized profiles illustrating the results of the de- and reprojection of the SZ emission of Abell 2142. Black: profile of the initial data maps; Red: profile obtained by de- and reprojecting the data maps. These profiles correspond to the azimuthal averaged of the maps on the left and on the right of Fig.~\ref{fig:MapSZ_A2142_deproj}, respectively. Bottom: relative residuals computed as in Fig.\ref{fig:sansvor_deproj}.}

\label{fig:SZ_A2142_deproj}
\end{center}
\end{figure}

We produced 100 maps from this initial map by adding simulated noise, assumed to be Gaussian, correlated, and inhomogeneous.
The result of the de- and reprojection of the SZ signal is shown in Fig.~\ref{fig:MapSZ_A2142_deproj}, for the averaged maps and in Fig.~\ref{fig:SZ_A2142_deproj}, for the azimuthally averaged profiles. In these figures, to help the comparison between the result of the de- and reprojection and the input data, we show the result of the de- and reprojection after convolution with \textit{Planck} PSF. The \textit{Planck} PSF is modelled here as a Gaussian of 7.1 arcmin FWHM and thus affects the region limited to 400 kpc around the cluster center.

\subsection{Reconstruction of the 2D gravitational potential of Abell 2142}

We here reconstruct the projected potential of the cluster from the X-ray (Eq.~(\ref{eq:phi_x})) and from the SZ data (Eq.~(\ref{eq:phi_sz})) for a polytropic index of $1.2$. This results from the fit ($\gamma=1.2\pm0.01$) of the density and pressure profiles of Abell 2142 \citep{tchernin16a} assuming a polytropic stratification of the intracluster gas (Eq.~(6)). This value is also in the range of polytropic indices expected in observations \citep[e.g.,][]{eckert13} and simulations \citep[e.g.,][]{tozzi01}. We discuss the validity of the polytropic relation in the discussion section below.

\begin{figure}
\begin{center}
  \includegraphics[height=0.7\columnwidth,angle=0]{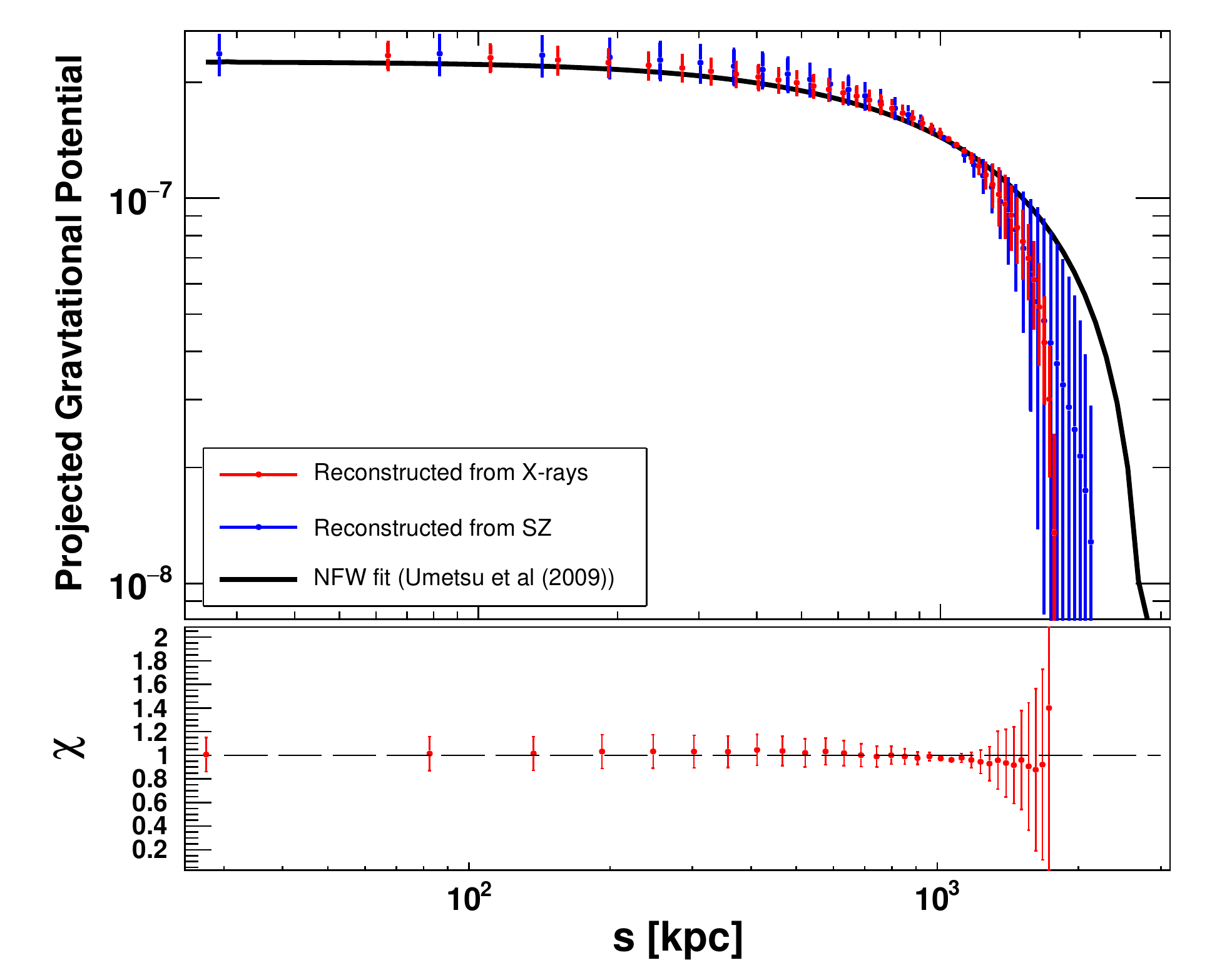}
\caption{Top: Normalized (over 1500kpc), azimuthally averaged, projected gravitational potential reconstructed from the X-ray (Eq.~(\ref{eq:phi_x})) and from the SZ signal (Eq.~(\ref{eq:phi_sz})) of Abell 2142 compared to the NFW fit of the lensing data performed by \citet{umetsu09}. Blue: projected potential reconstructed from the SZ signal; Red: projected potential reconstructed from the X-ray signal; Black: Fit of \citet{umetsu09}. Bottom: residuals between the potentials reconstructed from the X-ray and SZ signal computed with $(f_{SZ}(s)/f_{XR}(s))$ and the uncertainties on this quantity are obtained by error propagation. The dashed line represents the case where $f_{SZ}(s)=f_{XR}(s)$.}\label{fig:A2142_lensing}
\end{center}
\end{figure}

The profiles resulting from the azimuthal average of the projected gravitational potential reconstructed from X-ray and SZ measurements are shown in the top panel of Fig.~\ref{fig:A2142_lensing}. For comparison, we also plot the result of the NFW \citep{NFW} fit performed by \citet{umetsu09} based on the observed lensing signal. The bottom panel shows the residuals between the two reconstructions, computed as $(f_{SZ}(s)/f_{XR}(s))$ and the uncertainties on this quantity are obtained by error propagation.

Since the SZ observations are taken on an area slightly larger than that of the X-ray observations (as shown in Fig.~\ref{fig:A2142_data}), the potential recovered from the SZ signal extends to slightly larger cluster-centric radii than the potential recovered from the X-ray signal. Therefore, we have normalized the profiles in Fig.~\ref{fig:A2142_lensing} to the radial range [0,~ 1.5]~Mpc for the comparison. The two reconstructions are consistent with the profile derived from the NFW parameters of the fit to the gravitational-lensing observations by \citet{umetsu09} in the regions covered by the data. 
\\\\In the following section we show how the potentials recovered from the different observables can be jointly used to constrain the cluster potential despite the different fields-of-view (which were kept the same in this section only to illustrate the robustness of the reconstruction method, as mentioned in Sect.~2.1.2).

\section{Joint potential reconstruction}\label{sec:join}
\subsection {Method for the joint reconstruction}\label{sec:cov_computation}
Based on the gravitational potentials individually reconstructed from the X-ray and SZ measurements, we can now show the method for our joint potential reconstruction. This section aims at outlining the method for the joint reconstruction, the results and related discussions are reported in an accompanying study \citep[][in prep]{huber18}. The joint reconstruction method rests on the implementation of a general minimization procedure within the SaWLens framework \citep{merten09,merten14}. In its present state, the minimization implemented in the SaWLens framework allows us to recover the projected gravitational potential from both strong and weak-lensing measurements, using a $\chi^2$ minimization.  
The aim of the present section is to outline the method used to further constrain the projected potential by using additional cluster observables, namely the SZ effect, the kinematics of member galaxies and the X-ray signal.
The fit converges once the lensing observables, that are a combination of second derivatives of the projected gravitational potential \citep[see e.g.,][]{bart01}, have been reproduced successfully by the jointly reconstructed potential.

Relying on the assumption that all observables are measured independently, a $\chi^2$ minimization for the joint reconstructed potential $\psi$ is performed as follows: 

\begin{eqnarray}\label{eq:chi2total}
 \chi^2_\mathrm{total}\left(\psi\right) &=& \chi^2_\mathrm{weak~lensing}\left(\psi\right)+\chi^2_\mathrm{strong~lensing}\left(\psi\right) +\chi^2_\mathrm{X-ray}\left(\psi\right) \nonumber\\  &~&  + \chi^2_\mathrm{SZ}\left(\psi\right) +\chi^2_\mathrm{kinematics}\left(\psi\right)+Reg\left(\psi\right).
\end{eqnarray}
As the contributions based on gravitational lensing and on the regularization term ($Reg$) are described in detail in \citet{merten14}, in the following we focus on the contributions from SZ, X-ray, and galaxy kinematics. 

\begin{figure}
\begin{center}
  \includegraphics[height=0.6\columnwidth,angle=0]{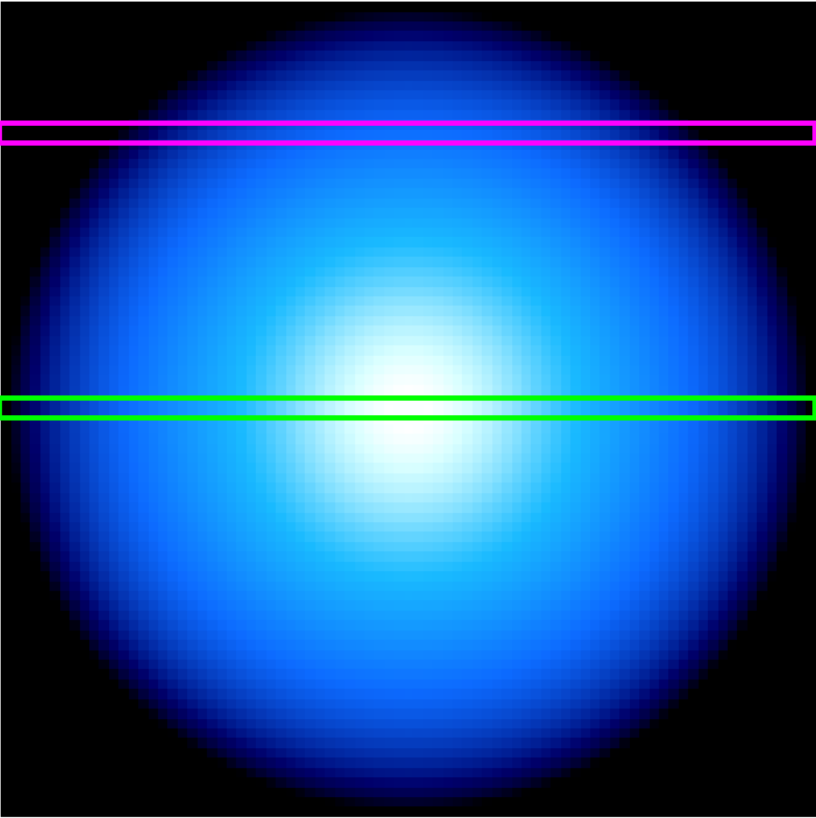}
\caption{Map representing the result of the average projected potential of Abell 2142 reconstructed from its X-ray measurements in Sect.~\ref{sec:recons_A2142}. The green (magenta) box indicate the pixels whose variance are enclosed in the diagonal entries of the block matrices with the corresponding color in Figs.~\ref{fig:A2142_low_high_Lfix_Lrun_total}-\ref{fig:A2142_low_high_Lfix_Lrun_edge}.}\label{fig:beta_covmatrix}
\end{center}
\end{figure}

At each pixel of the reconstructed grid, the contribution of the individually reconstructed potentials to the total  $\chi^2$  can be written as 

\begin{equation}\label{eq:chi_i}
\chi_i^2 = \left(A_i \bar\psi_i -\psi \right)^\mathrm{T} C^{-1}_i \left(A_i \bar\psi_i -\psi \right),
\end{equation}
where $C_i$ is the covariance matrix and where each potential $\psi_i$ ($i \in \lbrace \text{SZ, X-ray, kinematics}\rbrace$) has been individually reconstructed from the corresponding cluster observables (see Sect. 3 for the reconstruction based on SZ and X-ray observations, and \citet{sarli14}, for the reconstruction based on the velocity dispersion of the galaxy members). The scaling factor $A_i$ has been introduced because the reconstructed potential ($\psi_i$) is normalized in the reconstruction procedure. The optimization of this scaling factor and the minimization of the $\chi^2_i$ contribution will be described in detail in \citet[][in prep]{huber18}.
\\\\The covariance matrix of a data map of dimensions $n\times n$ has the dimension $n^2\times n^2$. Each entry of this $n^2\times n^2$ matrix can be computed as
\begin{equation}\label{eq:cov_mat}
  C_i(\vec{x},\vec{x'})=\langle \,(\psi_i(\vec{x})-\langle \psi_i(\vec{x})\rangle)\cdot (\psi_i(\vec{x'})\ -\langle \psi_i(\vec{x'})\rangle )\, \rangle\;,
\end{equation}
where $\langle \psi_i(\vec{x})\rangle$ denotes the mean of the value of the projected potential $\psi_i$ over a given number of realizations at the position $\vec{x}$. $\langle \psi_i(\vec{x})\rangle$ is illustrated for instance in Fig.~\ref{fig:beta_covmatrix}, where $\vec{x}$ represents any position on this map. The correlation between the scatter at any position $\vec{x}$ and $\vec{x'}$ of this map is contained in the covariance matrix computed with Eq.~(\ref{eq:cov_mat}). In case of data without fluctuations, we expect to obtain  $\psi_i(\vec{x})=\langle \psi_i(\vec{x})\rangle$ for all $\vec{x}$ and thus $C(\vec{x},\vec{x'})=0$. For $\vec{x}=\vec{x'}$, $C(\vec{x},\vec{x'})=\sigma^2_x$, implying that the diagonal values of the covariance matrix are equal to the variance of the data points.

The application of this method on a simulated cluster and the related results will be described in \citet[][in prep]{huber18}.

\subsection {Covariance matrix of the recovered projected gravitational potential}\label{sec:cov_app}
As shown in the previous sections, the gravitational potential's reconstruction from SZ, X-ray, and kinematics implies a deprojection of the data, which introduces correlations between the pixels of the reconstructed potential. Depending on the assumed geometry of the cluster, this correlation will create patterns in the covariance matrix $C_i$ (Eq.~(\ref{eq:cov_mat})).
Let us now study the correlations introduced by the R-L deprojection procedure during the reconstruction of the gravitational potential of a galaxy cluster from its X-ray measurements. 

To avoid overfitting the data, \citet{lucy94} suggested to add a regularization term to the fitting procedure performed in the R-L algorithm. This additional term is characterized by two parameters: the strength of the regularization $\alpha$, which calibrates the importance of this regularization term with respect to the data; and the smoothing scale $L$, which enters into the regularization term via 
\begin{equation}\label{eq:L}
  P(r|r')\sim \exp\left(-\frac{(r-r')^2}{L^2}\right)\; ,
\end{equation}
\citep[see][for details]{konrad13}. 
We expect both the smoothing scale $L$ and the resolution of the deprojected grid to induce correlations between pixels.

We investigate here how the resolution of the deprojected grid can affect the correlation between pixels, and analyze the correlations introduced by the smoothing parameter $L$, set in the deprojection method. For this purpose, we study the covariance matrix of the projected potential reconstructed from the X-ray observations of the cluster Abell 2142 (reconstructed in Sect. \ref{sec:recons_A2142}). The covariance matrix for the 100 reconstructed potentials has been derived from Eq.~(\ref{eq:cov_mat}) for two different grid resolutions:
\begin{enumerate}
\item{}\textit{High-resolution deprojected grid (HRG)}: the pixels of the grid used for the deprojection have the same size as the pixels of the input data grid (which is of dimension $100\times 100$ in our case, see Sect. \ref{sec:recons_A2142}).\\
\item{}\textit{Low-resolution deprojected grid (LRG)}: the deprojected grid has been arbitrarily chosen to contain $20\times 20$ pixels. One pixel of the deprojected grid contains $25$ pixels of the input data grid.
\end{enumerate}

\begin{figure}
\begin{center}
  \includegraphics[height=\columnwidth,angle=0]{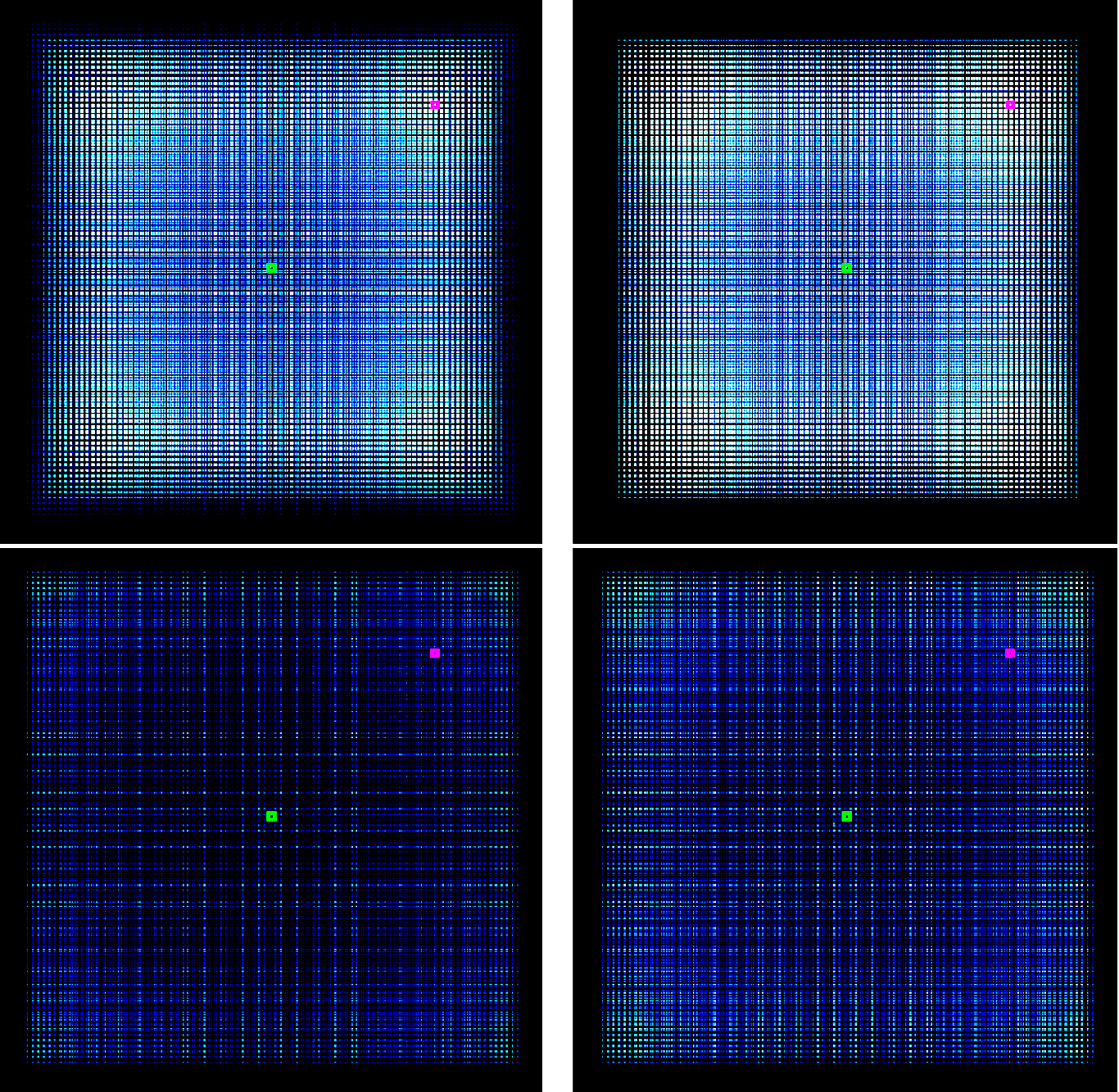}
\caption{Covariance matrices of the projected gravitational potential of the cluster Abell 2142 reconstructed from its X-ray observations (Sect.~\ref{sec:recons_A2142}), computed with the Eq.~(\ref{eq:cov_mat}). An illustration of a reconstructed potential is shown in Fig.~\ref{fig:beta_covmatrix}, with two selected lines of cells: one in magenta, and one in green. The information on the variance in these lines is contained in the diagonal entries of the block matrices with the matching color in the present figure and in Figs.~\ref{fig:A2142_low_high_Lfix_Lrun_middle} and \ref{fig:A2142_low_high_Lfix_Lrun_edge}.  The results for the HGR are shown in the top panels. Left: smoothing parameter fixed to $L=1$; right: smoothing parameter linearly increasing from $L=1$ to $L=10$. LRG results are shown on the bottom panels. Left: smoothing parameter fixed at $L=0.2$; right: smoothing parameter linearly increasing from $L=0.2$ to $L=2$. The same color scale has been used for the four panels, which runs from 0 (in black) to 0.004 (in white).}
\label{fig:A2142_low_high_Lfix_Lrun_total}
\end{center}
\end{figure}

\begin{figure}
\begin{center}
  \includegraphics[height=0.6\columnwidth,angle=0]{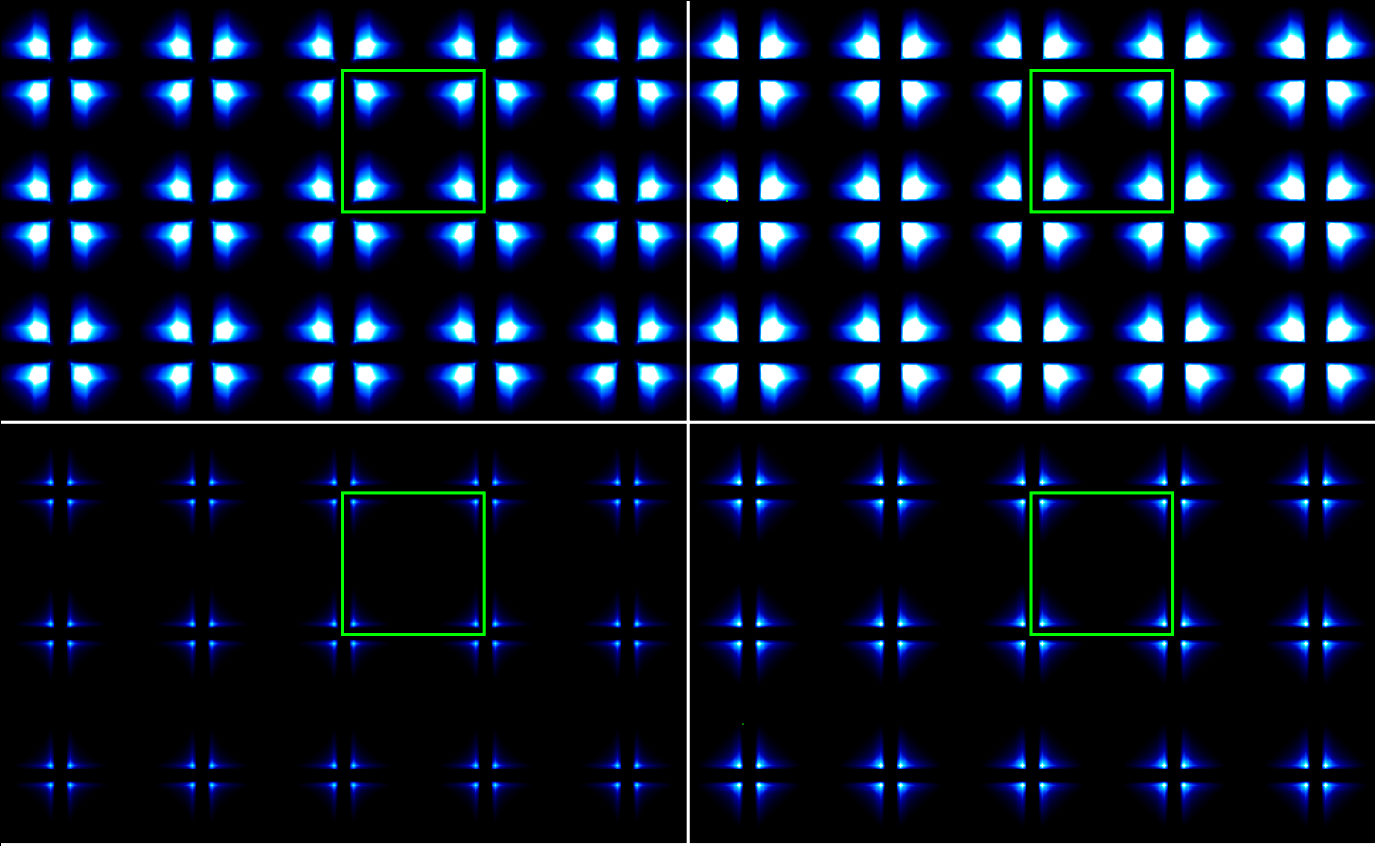}
\caption{Zoom into the central part of Fig.~\ref{fig:A2142_low_high_Lfix_Lrun_total}. The diagonal values of the enclosed block matrix contain the variance of the cells indicated by the green rectangle in Fig.~\ref{fig:beta_covmatrix}. Same color bars as in Fig.~\ref{fig:A2142_low_high_Lfix_Lrun_total}: From black to white indicating small to large fluctuations.}\label{fig:A2142_low_high_Lfix_Lrun_middle}
\end{center}
\end{figure}

\begin{figure}
\begin{center}
  \includegraphics[height=0.7\columnwidth,angle=0]{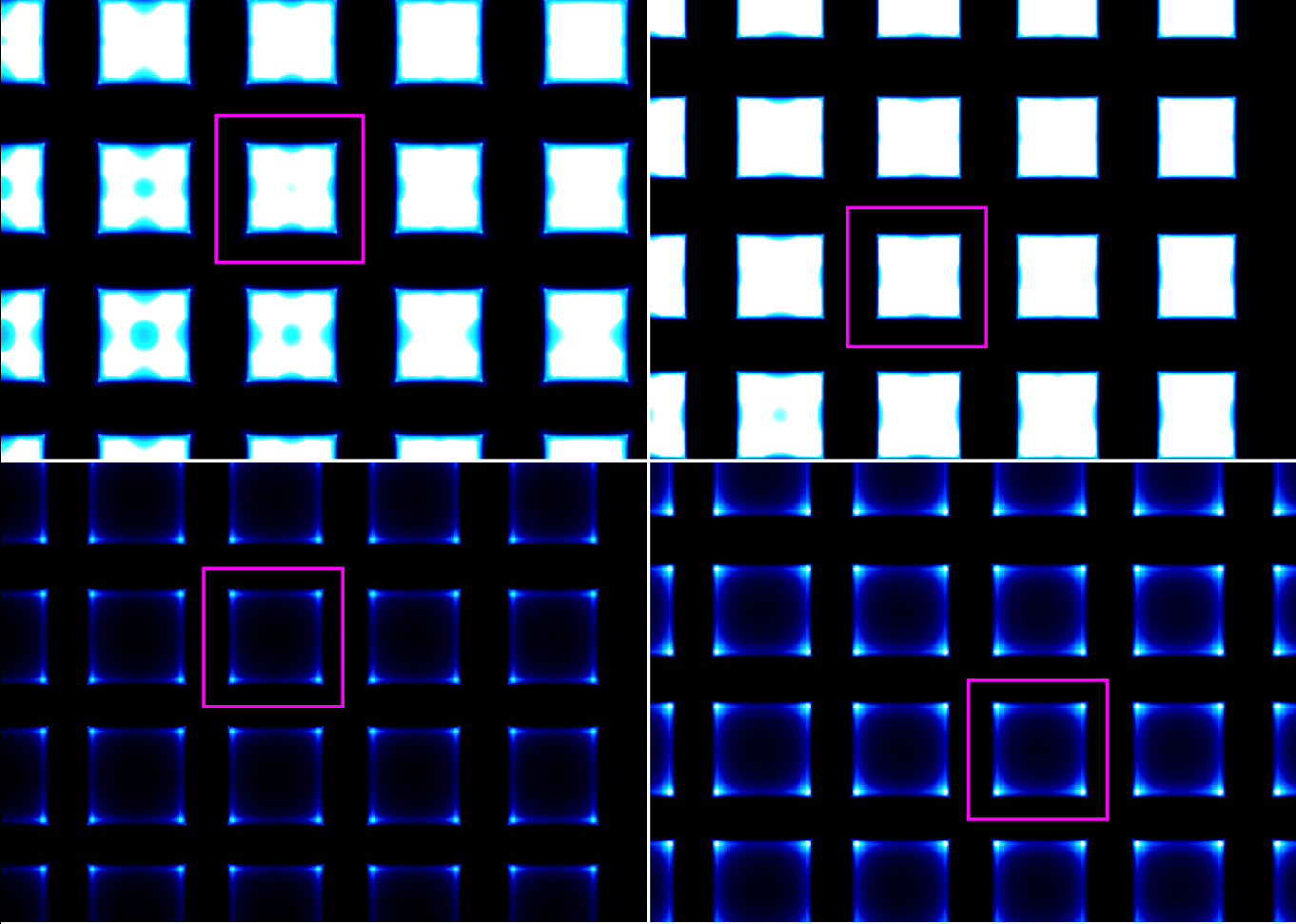}
\caption{Zoom into the corner of Fig.~\ref{fig:A2142_low_high_Lfix_Lrun_total}. The diagonal values of the enclosed block matrix contain the variance of the cells indicated by the magenta rectangle in Fig.~\ref{fig:beta_covmatrix}. Same color bars as in Fig.~\ref{fig:A2142_low_high_Lfix_Lrun_total}: From black to white indicating small to large fluctuations.}\label{fig:A2142_low_high_Lfix_Lrun_edge}
\end{center}
\end{figure}

In Figs.~\ref{fig:A2142_low_high_Lfix_Lrun_total}-~\ref{fig:A2142_low_high_Lfix_Lrun_edge}, we show the results the covariance matrices in the high- and low-resolution cases for a fixed smoothing scale corresponding to the smallest distance between two pixels of the deprojected grid ($L=1$ and $L=0.2$, for the HRG and the LRG, respectively) and for a smoothing scale linearily increasing with radius from 1 to 10 for the HRG and from 0.2 to 2 for the LRG.

We discuss the effects of $L$ and of the grid resolution in the discussion section.

\section {Discussion}\label{sec:discussion}

We have seen that the reconstruction method used in this analysis allows us to recover the 2D gravitational potential of galaxy clusters from X-ray and SZ observations.

Let us now discuss the quality of the potential reconstructions (Sect.~\ref{sec:recons_disc}), the validity of the assumptions made within the reconstruction method (Sect.~\ref{sec:ass_disc}), and how this information needs to be taken into account by the covariance matrix in the joint reconstruction  (Sect.~\ref{sec:covmatrix_disc}). For these last points, we focus our discussion on the cluster Abell 2142, which is our realistic case. 
We will discuss these points at a qualitative level, a full quantitative analysis of the effect of these assumptions on the potential reconstruction of a simulated cluster can be found in \citet[][in prep]{tchernin18}.

\subsection{General statements about the potential reconstruction method}\label{sec:recons_disc}

\subsubsection{Note on the quality of the reconstruction}
We showed the result of the deprojection of  $\beta$-profiles for three different cases of binning in Figs.~\ref{fig:mapsansvor_deproj}-\ref{fig:vor300_deproj}: without Voronoi tessellation, and with a Voronoi tessellation of 30 counts/bin and 300 counts/bin.

As expected from the deprojection procedure (see Sect.\ref{sec:methodRL}), we lose the information on the signal maps in the field corners and the signal is poorly reproduced at the cluster center ($<$2 pixels).
We observed that there is a tendency for the very last bin of the deprojected profile to fall below the input data. This artifact is due to the normalization of the spherical kernel (Eq.~(\ref{eq:sphk})), which, due to the pixelization of the grid, becomes too large at the last radial bin. Indeed, given that $s_1$ and $s_2$ are defined at the center of each pixel, the area of the grid that satisfies the condition $s_1^2+s_2^2<r^2$ is actually sightly larger than the expected value of $2\pi r^2$ for the last radial bin.
Nevertheless, we do not expect this effect to influence our potential reconstruction significantly.

\begin{figure}
\begin{center}
  \includegraphics[height=\columnwidth,angle=270]{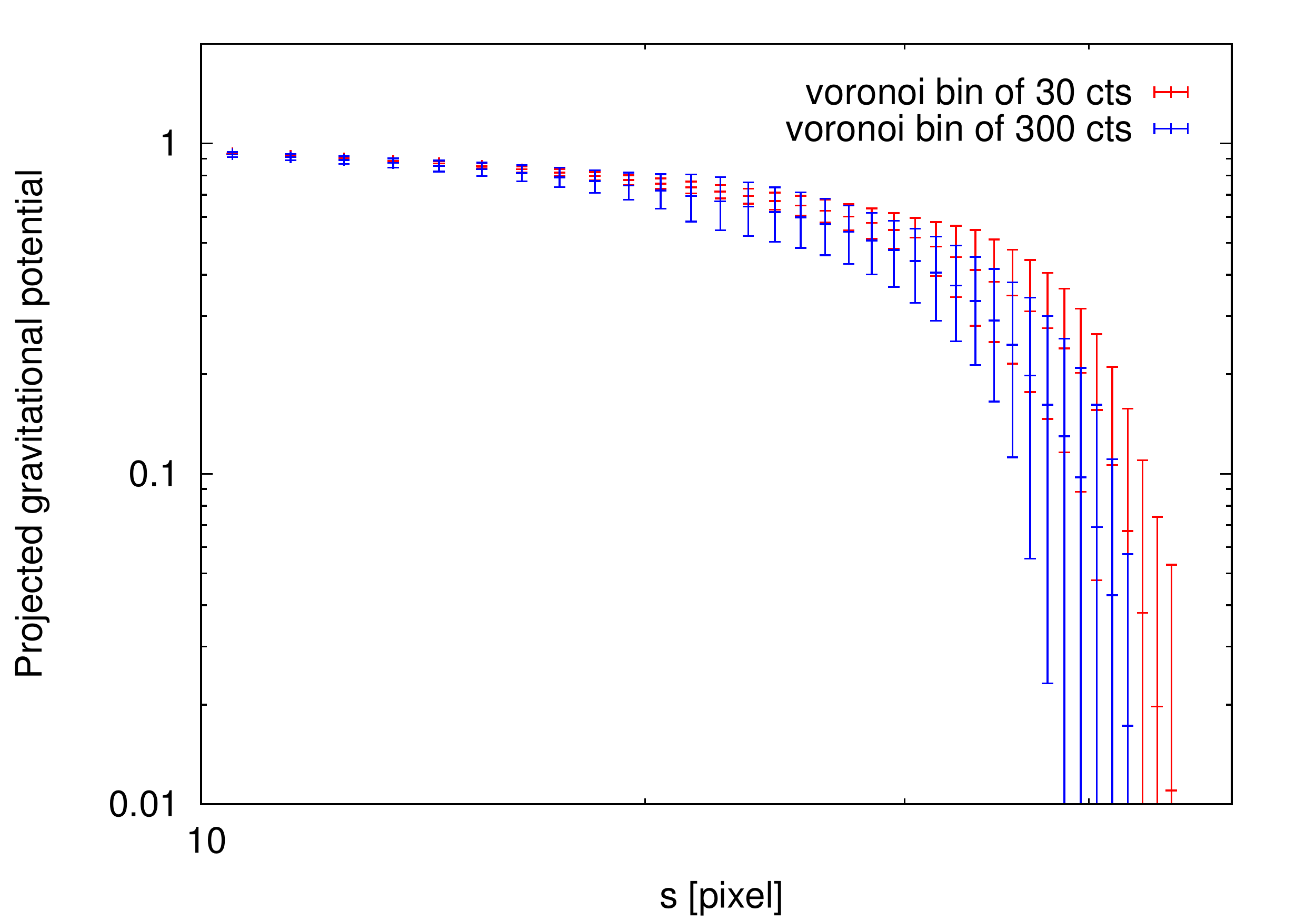}
\caption{Azimuthally averaged projected gravitational potential profiles recovered from the simulated $\beta$-profile.  Red: from a Voronoi-tessellated map of 30 counts/bin; Blue: from a Voronoi-tessellated map of 300 counts/bin. Here, it has been assumed that the $\beta$-profile represents the X-ray surface brightness. The potential profiles have been reconstructed using Eq.~(\ref{eq:phi_x}).}\label{fig:lensing_vor300_30_withoutvor}
\end{center}
\end{figure}

We also investigated the effect of the smoothing at large cluster-centric radii produced by the Voronoi tessellation on the reconstructed potential and showed that the resulting reconstructed projected potential profiles have large error bars at large radii. This is illustrated in Fig.~\ref{fig:lensing_vor300_30_withoutvor}, where we assumed that the $\beta$-profile emission was representing X-ray data \citep[motivated by e.g.,][]{jones84}. 
As stated before, this effect is nevertheless not expected to have any impact in the joint reconstructed potential, as the individual potential values with large error bars are down-weighted in the combined reconstruction.

\subsubsection{Note on SZ and X-ray reconstruction of the projected potential of the simulated g1-cluster}
In Sect.~\ref{sec:recons_g1} we recovered the projected gravitational potential of the simulated g1-cluster from X-ray and SZ mock data. Both reconstructed potentials are consistent with the true projected potential (see Fig.~\ref{fig:G1_lensing}). We observed a radial trend between the reconstructed and the true projected gravitational potentials, which may indicate that the cluster is in a more complicated physical state than the assumed equilibrium. Such a study can thus be used to assess the physical state within the cluster, as it was pointed out for instance in \citet{tchernin15} for the cluster A1689. On the other hand, assuming that the equilibrium assumptions are valid, a joint analysis could be used to constrain the geometry of the cluster \citep[see e.g.,][]{filippis05}.
For instance, joint analyses have successfully been applied to constrain the triaxiality in several galaxy clusters \citep[e.g.,][]{morandi11, morandi12a, morandi12b} using a parametric method. The comparison of the constraints obtained from these analyses with our method is beyond the scope of our paper. A comprehensive study of the joint analysis and of the related systematics will be described in \citet[][in prep]{huber18}.

\subsubsection{Note on SZ and X-ray reconstruction of the projected potential of the cluster Abell 2142}
We reconstructed the projected gravitational potential of Abell 2142 from SZ and X-ray measurements in Sect.~\ref{sec:recons_A2142} and obtained consistent results for the two reconstructed potentials (see Fig.~\ref{fig:A2142_lensing}). Given that the uncertainties of the NFW parameters were not taken into account in the lensing potential estimated from \citet{umetsu09} and that the region outside $\sim$2000 kpc is  not testable with the data used in this analysis, any conclusion about the state of the gas in this region would be speculative (see Sect.~7.2.2).
\subsection{Validity of the assumptions for Abell 2142}\label{sec:ass_disc}
\subsubsection{The geometry of the cluster}
\label{sec:sym}
The assumption of intrinsic spherical cluster symmetry is quite restrictive. The result of the de- and reprojection shows that a large part of the information contained in the data is lost in the implied averaging (see for instance Figs.~\ref{fig:MapXR_A2142_deproj} and \ref{fig:MapSZ_A2142_deproj}). Nevertheless, we have shown that this strong symmetry assumption still allows us to satisfactorily reconstruct potentials for the specific cases of Abell 2142 (as well as for the g1-cluster). The effect of this assumption on the reconstruction of the potential of more disrupted clusters is under investigation \citep[][in prep.]{tchernin18}.

Furthermore, the generalization to spheroidal symmetry is ongoing \citep{reblinsky00,puchwein06} and has already been successfully tested in \citet{majer16} for the reconstruction from SZ data. This generalization is crucial, as we aim at recovering a joint projected gravitational potential on a 2D map. However, even spheroidal symmetry will introduce correlations between the pixels that should be carefully taken into account in the joint analysis. 

\subsubsection{The equilibrium assumptions}\label{sec:HE_disc}
The gravitational potential reconstruction from X-ray, SZ and kinematics is based on the~hydrostatic equilibrium and the polytropic stratification assumptions. Those strong assumptions are not expected to be valid accross the entire observable range of cluster-centric radii. We want to discuss here how this may affect the joint reconstruction of the potential of Abell 2142.

\begin{figure}
\begin{center}
  \includegraphics[height=0.6\columnwidth,angle=0]{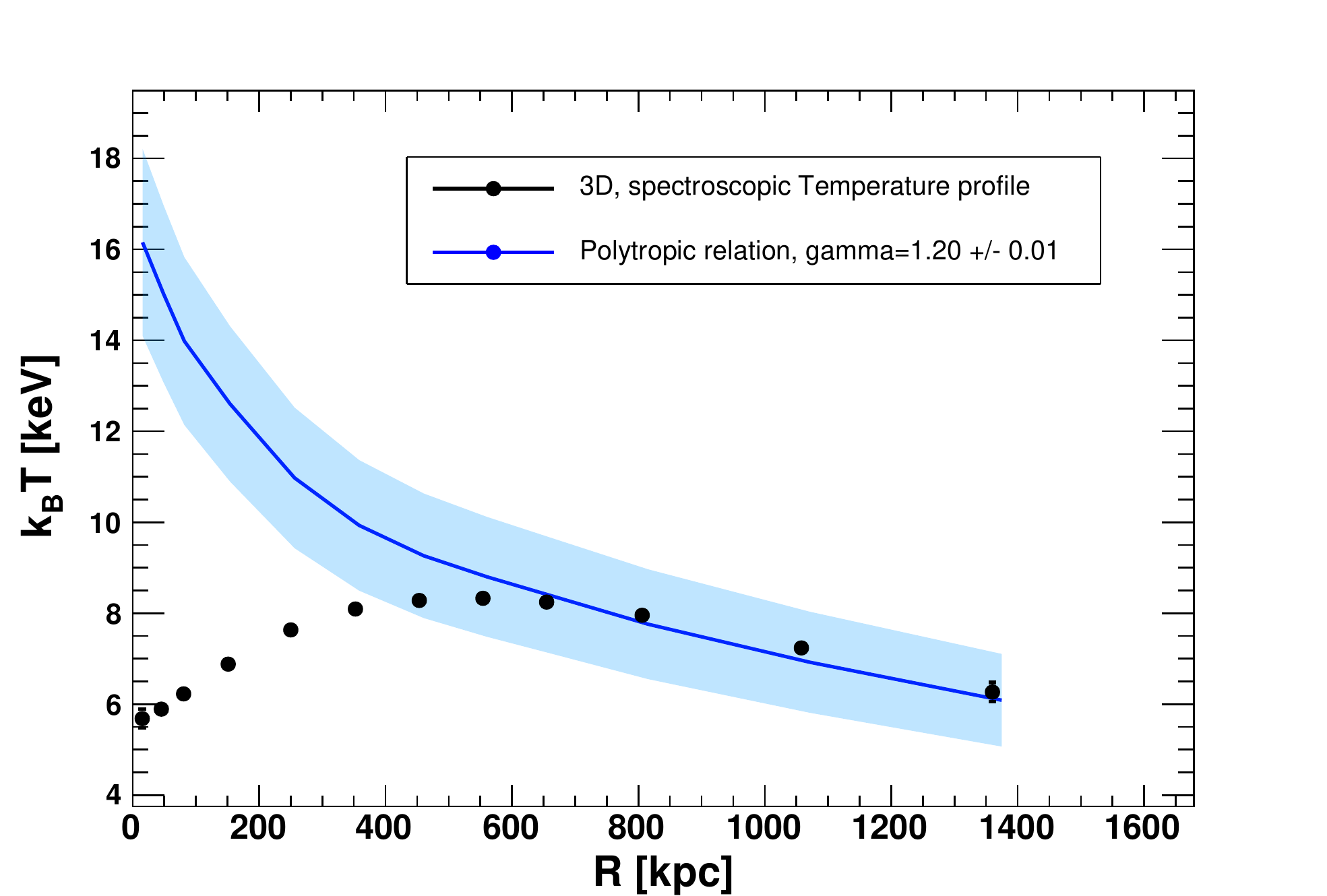}
\caption{Comparison of the deprojected temperature profile from Abell 2142 with the temperature profile derived assuming polytropic stratification and a polytropic index of $\gamma=1.20\pm0.01$. Both the deprojected spectroscopic temperature profile and the spectroscopic density profile are taken from \citet{tchernin16a}.}\label{fig:T_poly}
\end{center}
\end{figure}

\begin{enumerate}
\item{}Due to mixing of the ICM with the infalling material from the large-scale structure, simulations tend to show that the equilibrium assumptions are not valid in the cluster outskirts \citep[see, e.g.,][]{reiprich13,nagai11,vazza13,zhuravleva13}.
In the case of Abell 2142, the gas in the cluster seems to be in hydrostatic equilibrium with the gravitational potential in the region from 400kpc out to about 3Mpc ($\sim$ $2R_{500}$). This region was tested in \citet{tchernin16a} based on the same measurements as those used here and no evidence for a deviation from the hydrostatic equilibrium was found \citep[see e.g., the recovered hydrostatic mass of the cluster in Fig. 13 of][]{tchernin16a}. The region inside 400 kpc was removed from the analysis due to the large PSF of the \textit{Planck} instrument. \\

\item{}Based on the spectroscopic analysis of the \textit{XMM-Newton} measurements performed in \citet{tchernin16a}, the temperature profile increases out to 500kpc and then decreases with the cluster-centric radius \citep[see Fig. 3 in][]{tchernin16a}. This implies that the temperature in the central region cannot be approximated by a polytropic stratification. Thus, the polytropic stratification is not valid in the region inside 500kpc but valid at larger radii, as shown in Fig.~\ref{fig:T_poly}. The polytropic exponent has been set here to agree with a fit of the density and pressure profiles assuming polytropic stratification.

\end{enumerate}
The gravitational potential derived from lensing observations has been recovered without requiring equilibrium assumptions \citep[see e.g.,][for a review]{bart01}. Therefore the comparison of the lensing potential with the potentials reconstructed from X-ray and SZ observations carries interesting information about the physical state of the cluster. As the R-L deprojection algorithm is ``local", in the sense that the regions at different cluster-centric radii are not mixed in the deprojection, this comparison allows us to determine the regions where the assumptions made do not hold. Such regions can be removed from our analysis or can be down-weighted in the joint reconstruction by adapting the covariance matrix, as discussed below. In the present case, the potentials recovered from X-ray, SZ and lensing are compared in Fig. 20. As we can see, all potentials are consistent. This is unexpected at the cluster center, as the polytropic assumption is not valid there. However, the potential being smoothed by the projection, this may hide the expected discrepancy near the center. Nevertheless, in the joint reconstruction, the central region of the reconstructed potentials needs to be down-weighted  with respect to the lensing potential, as described in the following section. 

\subsection{Covariance matrix}\label{sec:covmatrix_disc}
The covariance matrix is one of the most important quantities for our joint reconstruction, as it contains the information about the quality of each individual reconstruction, pixel per pixel. Here we want  to discuss how the different assumptions and the technical details of the observations (resolution of the grid, deprojection of the data) affect the overall shape of the covariance matrix. The reconstruction requires the deprojection of the data and, as we have seen in Sect.~\ref{sec:cov_app}, the assumption on the geometry of the cluster, the smoothing scale, and the resolution of the deprojected grid affect the covariance matrix.

\begin{itemize}
\item{\textit{Effect of the smoothing scale on the correlation between pixels}}:
\\\\
The effect of the smoothing scale is two-fold: On one hand, it introduces correlations between pixels; on the other, it reduces fluctuations (which at the same time reduce the correlations, see Eq.~(\ref{eq:cov_mat})). Thus, these two effects act in opposite directions, and the relative value of $L$ with respect to the dimension of the grid determines which effect dominates. To test this statement, we kept the value of $L$ fixed at the distance between two neighboring pixels (to simulate the case with negligible correlation), and computed the covariance matrices. We then repeated this exercise for a smoothing parameter varying linearly with radius from the distance between two neighboring pixels to one tenth of the grid dimension. This is illustrated in Figs.~\ref{fig:A2142_low_high_Lfix_Lrun_total}-\ref{fig:A2142_low_high_Lfix_Lrun_edge}. The comparison of the left and right panels of Figs.~\ref{fig:A2142_low_high_Lfix_Lrun_total}-\ref{fig:A2142_low_high_Lfix_Lrun_edge} clearly shows that the covariance matrix reaches higher values when $L$ is large. We also checked that this trend changes when we continue to increase the value of $L$. Indeed, for a value of $L$ comparable for instance to the half of the grid dimension, the correlation between pixels is reduced. This is due to the fact that this smoothing procedure acts like an averaging process: If $L$ is as large as half of the dimension of the map, the fluctuations expected at large distance from the cluster center will be less pronounced and the value of the factor $(x(i)\ -\langle x(i)\rangle)$ in Eq.~(\ref{eq:cov_mat}) will be small. \\

\item{\textit{Effect of the grid resolution on the correlation between pixels}}:
\\\\
The resolution of the deprojected grid has a considerable effect on the covariance matrix. This becomes obvious comparing the top (HRG) and bottom (LRG) panels of Figs.~\ref{fig:A2142_low_high_Lfix_Lrun_total}-\ref{fig:A2142_low_high_Lfix_Lrun_edge}. In the cases shown, a pixel of the LRG grid bundles 25 pixels of the HRG grid. The decrease of the correlation between pixels when passing from HRG to LRG is expected on the one hand because the statistics per pixel improves, and on the other hand, because the value at each pixel in the LRG corresponds to the average over 25 pixels of the HRG, both effects resulting in the decrease of the fluctuations in the LRG.\\

\item{\textit{Some considerations about the shape of the block matrices contained in the covariance matrix}}:
\\\\
We note that all four covariance matrices in Fig.~\ref{fig:A2142_low_high_Lfix_Lrun_total} have smaller values in the center than in the corners. It is indeed expected that the covariance matrix elements are largest where the projected potential is less constrained by the data, that is,\ at large distance from the center and in the corners, where the information about the initial data map is lost due to the deprojection procedure (see Sect. 2.1 for a detailed description of the deprojection method). The overall shape of the covariance matrix can be understood considering its diagonal elements. For instance, we indicated in Fig.~\ref{fig:A2142_low_high_Lfix_Lrun_middle} the block matrix whose diagonal elements correspond to the variance of the potential values in the region delimited by the same color in Fig.~\ref{fig:beta_covmatrix}. As shown there, the correlations decrease towards the center, where the potential values are largest, while the correlations increase in the cluster outskirts, where the uncertainties in the reconstructed potential are largest.\\

\item{\textit{Effect of the assumed symmetry on the correlation between pixels}}:
\\\\
As the R-L method allows the deprojection of each line of sight individually, we could expect the deprojected data points to be uncorrelated. However, this is not what we observed. Indeed, the assumption of spherical symmetry introduces correlations between all values having the same projected radius. Therefore, the nondiagonal entries of the covariance matrix contain nonzero values.

Since the deprojection is the first step of the potential reconstruction, we expect to see similar effects in the covariance matrices derived from the SZ signal (see for instance Eq.~(1) and Sect. 3). The deprojection kernel in the case of the kinematics data is slightly different from the spherical kernel used for X-ray and SZ reconstruction \citep{sarli14}. Therefore, the covariance matrix from kinematic mesurements may be different. This study is ongoing and will be described in a follow-up paper.\\

\item{\textit{Effect of the assumed physical state of the cluster on the correlation between pixels}}:
\\\\
The effects of deviations from equilibrium assumptions on the potential reconstruction cannot be computed with Eq.(\ref{eq:cov_mat}), but rather, it is an additional piece of information which needs to be added to the covariance matrix prior to the joint reconstruction. In the case of Abell 2142, the values of the covariance matrices for the potentials reconstructed from the SZ and X-ray measurements (with Eq.~(\ref{eq:cov_mat})) need to be adapted at the cluster center to take into account that the polytropic assumption is known not to be valid there. The overall modification of the covariance matrices needs to be done in a consistent way for each individual contribution $\chi^2_i$ in Eq.~(\ref{eq:chi2total}) to ensure the convergence of the minimization procedure toward a jointly constrained potential. This needs to be done carefully, once each single contribution to $\chi^2_\mathrm{total}$ is known.
\end{itemize}

\section {Conclusion}\label{sec:conclusion}

We generalized the R-L deprojection method to a grid. We tested this method for the first time on the potential reconstruction from SZ and X-ray data on a 2D grid. We showed that the gravitational potential reconstructions of individual clusters based on their X-ray, SZ and gravitational lensing observations are consistent. The quality of the reconstruction from each observable separately, contained in the covariance matrix of the reconstructed potential, is a key element of the joint reconstruction. We also showed how the assumption on the intrinsic cluster symmetry, the smoothing scale, and the resolution of the deprojected grid affect the covariance matrix.

The goal is to combine each of these contributions weighted by the quality of its own reconstruction. The inclusion of the contribution of each observable to the total $\chi^2$ minimization in the SaWLens framework is ongoing \citep[][in prep]{huber18}.

The results of the reconstruction of the potentials of the simulated g1-cluster and of the real Abell 2142 cluster are very encouraging. These reconstructions have been obtained assuming spherical symmetry, hydrostatic equilibrium, and polytropic stratification. A dedicated analysis of the effects of these assumptions on the reconstructed potential is studied in \citet[][in prep]{tchernin18}. We shall generalize this work to spheroidal symmetry \citep[e.g.,][]{puchwein06,reblinsky00,majer16} and we aim at dropping the polytropic stratification assumption to recover more realistic cluster potentials.

This study is part of a larger project whose main goal is to perform the joint potential reconstruction of galaxy clusters \citep[from the CLASH sample for instance,][]{postman12}, based on all cluster observables.

\begin{acknowledgements}
We thank Dr. Matteo Maturi for providing private codes used in this study. CT acknowledges the financial support from the Swiss National Science Foundation (P2GEP2\_159139). KH acknowledges support by the DFG cluster of excellence "Origin and Structure of the Universe". JM has received funding from the People Programme (Marie Curie Actions) of the European Unions Seventh Framework Programme (FP7/2007-2013) under REA grant agreement number 627288. MM acknowledges support from the Italian  Ministry of Foreign Affairs and International Cooperation, Directorate General for Country Promotion, from INAF via  PRIN-INAF 2014 1.05.01.94.02, and from ASI via contract ASI/INAF/I/023/12/0. This project was supported in part by the Baden-W\"urttemberg Foundation under project ``Internationale Spitzenforschung II/2'', by the Collaborative Research Centre TR-33 ``The Dark Universe'' as well as project BA 1369/17 of the Deutsche Forschungsgemeinschaft. We acknowledge partial support by the Germany-Israel GIF I-1341-303.7/2016 and DIP STE1869/2-1 GE625/17-1.
\end{acknowledgements}

\end{document}